\documentclass[aps,twocolumn,showpacs,superscriptaddress]{revtex4}
\usepackage{epsfig}
\usepackage{times}

\begin{document}

\title{Role-separating ordering in social dilemmas controlled by topological frustration}

\author{Marco A. Amaral}
\email{marcoantonio.amaral@gmail.com}
\affiliation{Departamento de F\'\i sica, Universidade Federal de Minas Gerais, Caixa Postal 702, CEP 30161-970, Belo Horizonte - MG, Brazil}

\author{Matja{\v z} Perc}
\affiliation{Faculty of Natural Sciences and Mathematics, University of Maribor, Koro{\v s}ka cesta 160, SI-2000 Maribor, Slovenia}
\affiliation{CAMTP -- Center for Applied Mathematics and Theoretical Physics, University of Maribor, Krekova 2, SI-2000 Maribor, Slovenia}

\author{Lucas Wardil}
\affiliation{Departamento de Fisica, Universidade Federal de Ouro Preto, Ouro Preto, MG, Brazil}

\author{Attila Szolnoki}
\affiliation{Institute of Technical Physics and Materials Science, Centre for Energy Research, Hungarian Academy of Sciences, Post Office Box 49, H-1525 Budapest, Hungary}

\author{Elton J. da Silva J{\'u}nior}
\affiliation{Departamento de F\'\i sica, Universidade Federal de Minas Gerais, Caixa Postal 702, CEP 30161-970, Belo Horizonte - MG, Brazil}

\author{Jafferson K. L. da Silva}
\affiliation{Departamento de F\'\i sica, Universidade Federal de Minas Gerais, Caixa Postal 702, CEP 30161-970, Belo Horizonte - MG, Brazil}

\begin{abstract}
``Three is a crowd'' is an old proverb that applies as much to social interactions, as it does to frustrated configurations in statistical physics models. Accordingly, social relations within a triangle deserve special attention. With this motivation, we explore the impact of topological frustration on the evolutionary dynamics of the snowdrift game on a triangular lattice. This topology provides an irreconcilable frustration, which prevents anti-coordination of competing strategies that would be needed for an optimal outcome of the game. By using different strategy updating protocols, we observe complex spatial patterns in dependence on payoff values that are reminiscent to a honeycomb-like organization, which helps to minimize the negative consequence of the topological frustration. We relate the emergence of these patterns to the microscopic dynamics of the evolutionary process, both by means of mean-field approximations and Monte Carlo simulations. For comparison, we also consider the same evolutionary dynamics on the square lattice, where of course the topological frustration is absent. However, with the deletion of diagonal links of the triangular lattice, we can gradually bridge the gap to the square lattice. Interestingly, in this case the level of cooperation in the system is a direct indicator of the level of topological frustration, thus providing a method to determine frustration levels in an arbitrary interaction network.
\end{abstract}

\maketitle

\section{Introduction}
\label{Introduction}
The evolution of cooperation is still a major open problem in biological and social sciences \cite{pennisi_s05}. After all, why should self-interested individuals incur costs to provide benefits to others? This puzzle has been traditionally studied by means of evolutionary game theory, and with remarkable success \cite{maynard_82, weibull_95, hofbauer_98, mestertong_01, nowak_06}. The prisoner's dilemma game \cite{axelrod_84, nowak_06}, for example, is the classical setup of a social dilemma. The population is best off if everybody would cooperate, but the individual does best if it defects, and that regardless of what other choose to do. In classical game theory, the Nash equilibrium of the prisoner's dilemma game, indeed the rational choice, is thus to defect. Nevertheless, cooperation flourishes in nature, and it is in fact much more common as could be anticipated based on the fundamental Darwinian premise that only the fittest survive. Humans, birds, ants, bees, and even different species between one another, all cooperate to a more or less great extent \cite{wilson_71, skutch_co61, nowak_11}.

An important step forward in understanding the evolution of cooperation theoretically was to consider spatially structured populations, modeled for example by a square lattice, which has been done first by Nowak and May \cite{nowak_n92b} who discovered network reciprocity. In spatially structured populations cooperators may survive because of the formation of compact clusters, where in the interior they are protected against the invasion of defectors. Other prominent mechanisms that support the evolution of cooperation include kin selection \cite{hamilton_wd_jtb64a}, mobility and dilution \cite{Alizon2008, Sicardi2009}, direct and indirect reciprocity \cite{trivers_qrb71, axelrod_s81}, network reciprocity \cite{nowak_n92b, santos_prl05, santos_pnas06, gomez-gardenes_prl07}, group selection \cite{wilson_ds_an77}, and population heterogeneity \cite{szolnoki_epl07, perc_pre08, santos_n08, santos_jtb12}. In particular, research in the realm of statistical physics has shown that properties of the interaction network can have far reaching consequences for the outcome of evolutionary social dilemmas \cite{zimmermann_pre04, zimmermann_pre05, fu_pre09, du_wb_epl09, lee_s_prl11, gomez-gardenes_epl11, ohdaira_jasss11, tanimoto_pre12, santos_md_srep14, pavlogiannis_srep15, wu_zx_epl15, hindersin_pcbi15, chen_w_pa16} (for reviews see \cite{szabo_pr07, roca_plr09, perc_bs10, perc_jrsi13, pacheco_plrev14, wang_z_epjb15, wang2015universal}), and moreover, that heterogeneity in general, be it introduced in the form of heterogeneous interaction networks, noisy disturbances to payoffs, or other player-specific properties like the teaching activity or the propensity to acquire new links over time, is a strong facilitator of cooperation \cite{vukov_pre06, perc_njp06a, tanimoto_pre07b, perc_pre08, szolnoki_epl08, szolnoki_epjb08, jiang_ll_pre09, devlin_pre09, shigaki_epl12, hauser_jtb14, yuan_wj_pone14, iwa_pha15, amaral_jpa15, tanimoto2015network, liu_rr_epl15, javarone_epjb16, amaral_pre16, chen_w_pa16, matsuzawa2016spatial, javarone2016role, javarone2016conformity}.

However, the impact of a structured population is not always favorable for the evolution of cooperation. If the interaction network links three individuals into a triangle, it may be challenging, or even impossible, to come up with a distribution of strategies that ensures everybody is best off (even if one assumes away the constrains of the evolutionary competition) \cite{wardil_epl09}. In the snowdrift game, anti-coordination of the two competing strategies is needed for an optimal outcome of the game. Clearly, in a triangle, if one individual cooperates and the other defects, the third player is frustrated because it is impossible to choose a strategy that would work best with both its neighbors. Similarly frustrated setups occur in traditional statistical physics, and have in fact been studied frequently in solid-state physics \cite{Robinson2011, binder_prb80}. In anti-ferromagnetic systems, for example, spins seek the opposite state of their neighbors, and again, it is clearly impossible to achieve this in a triangle. As noted above, the snowdrift game is in this regard conceptually identical, and thus one can draw on methods of statistical physics and on the knowledge from related systems in solid-state physics to successfully study the evolutionary dynamics of cooperation in settings that constitute a social dilemma.

The manifestation of topological frustration in the snowdrift game, however, can depend strongly on how the players update their strategies during the evolutionary process. In the light of recent human experiments \cite{gracia-lazaro_srep12, gracia-lazaro_pnas12, Grujic2014, vukov_njp12, blume_a_geb10, szabo_pr07, bonawitz_cg14}, we here consider not only the generally used imitation dynamics, but also the so-called logit rule (also known as myopic dynamics) \cite{blume_l_geb95, szabo_jtb12b, szabo_pre05}. The latter can be considered as more innovative, allowing players to choose strategies that are not within their neighborhood if they provide a good response to the strategies of their neighbors. Although the long term evolution in animals is best described by imitation dynamics, humans tend to be more inventive, and thus their behavior aptly described also by innovative dynamics \cite{wedekind_pnas96, Dalton2010, vukov_njp12, macy_pnas02, blume_a_geb10, bonawitz_cg14, nowak_06, szabo_pr07, roca_epjb09, sysiaho_epjb05, grujic_pone10, szolnoki_epl11}. Indeed, the impact of the logit rule and of closely related strategy updating protocols on the outcomes of evolutionary games on the square lattice has been studied extensively \cite{sysiaho_epjb05, roca_pre09, roca_epjb09, szabo_pre10, szabo_jtb12b, szabo_jtb12}, but there the topological frustration is absent.

In what follows, we fill this gap by studying the snowdrift game on the triangular lattice, as well as the transition from the square to the triangular lattice, both by means of mean-field approximations and Monte Carlo simulations. Our main objective is to reveal how an inherent topological frustration affects the evolutionary outcomes. We observe fascinating honeycomb-like patterns, and we devise an elegant method to determine the level of frustration in an arbitrary interaction network through the stationary level of cooperation. Before presenting the main results, we first describe the mathematical model, and we conclude with a discussion of the wider implications of our findings.

\section{Mathematical model}
\label{Model}
In our model, players have only two possible strategies, namely cooperation (C) and defection (D), and the game is played in a pairwise manner as defined by the interaction network. During each pairwise interaction players receive a payoff according to the payoff matrix \cite{szabo_pr07, nowak_06}
\begin{equation}\label{paymatrix}
 \bordermatrix{~ & C & D \cr
                  C & R & S \cr
                  D & T & P \cr},
\end{equation}
where $T\in[0,2]$, $S\in[-1,1]$ and $R=1$, $P=0$. This parametrization is useful as it spans four different classes of games, namely the prisoner's dilemma game (PD), the snowdrift game (SD), the stag-hunt game (SH), and the harmony game (HG) \cite{szabo_pr07, szabo_pre05, roca_pre09, wang_z_srep12}. After players collect their payoff, they may change their strategies based on a particular strategy updating rule. In this paper, we consider the logit rule and compare it with the classical imitation rule.

The logit rule is based on the kinetic Ising model of magnetism (also known as Glauber dynamics \cite{glauber_jmp63, binder_prb80}). The site will change its strategy with probability
\begin{equation}\label{isingeq}
p(\Delta u_{i})=\frac{1}{ 1+e^{-(u_{i*}-u_{i})/K} }
\end{equation}
where $u_i$ is the site current payoff, $u_{i*}$ is the site's payoff if it changed to the opposite strategy, and the states of neighborhood remain unchanged. Finally $K$ is a parameter that measures the irrationality of players. In the literature $K$ is usually set between $K\in[0.001,0.4]$ to simulate a small, but non-zero, chance of making mistakes \cite{szabo_pr07, nowak_06}, we set it to be $0.1$. Mathematically, the model is equivalent to the statistics used in physics to describe the dynamics of spins in a Fermi-Dirac distribution and is widely used in evolutionary dynamics \cite{hauert_ajp05}. In the context of game theory, this kind of update rule (also know as myopic best response \cite{szabo_jtb12b}) is regarded as a player asking himself what would be the benefits of changing his strategy (even when there is no neighbor with different strategy). This means that the logit rule is an innovative dynamic, since new strategies can spontaneously appear. Recently, the logit rule has been the focus of many works \cite{szabo_jtb12b, szabo_jtb12, szabo_pr07, szabo_pre10, roca_pre09, roca_epjb09, sysiaho_epjb05, gracia-lazaro_pnas12, gracia-lazaro_srep12, Grujic2014} as it leads to very different results compared to imitation models. As we see, this rule is closely related to rational analysis of a situation, instead of the reproduction of the ``fittest'' behavior. Although evolutionary game theory has its bases rooted in biological populations dynamics, recent works shows that the modeling of humans playing games can have more in common with innovative dynamics \cite{gracia-lazaro_pnas12, gracia-lazaro_srep12, Grujic2014, hauert_n04, roca_epjb09}.

The imitation rule, or imitation dynamics, is one of the most common update rules in iterated evolutionary game theory \cite{nowak_06, szabo_pr07}, and is based on the concept of the fittest strategy reproducing to neighboring sites. Here we will use it as a baseline for comparison with our results. Site $i$ will update its state by randomly choosing one of its neighbors, $j$, and then comparing their payoff. Site $i$ adopts the strategy of $j$ with probability
\begin{equation}\label{imiteq}
p(\Delta u_{ij})=\frac{1}{ 1+e^{-(u_{j}-u_{i})/K} },
\end{equation}
where $u_{i,j}$ is the total payoff of site $i,j$ \cite{szabo_pre07}. Note that player $i$ can only change its strategy to the ones available in its neighborhood. This means that new strategies can never appear once extinguished and players never ``explore'' new strategies, which can be interpreted as a non-innovative dynamic. This model is associated with biological processes, where each strategy is regarded as a specie, and once extinguished it will never re-appear \cite{szabo_pr07, nowak_06, nowak_s04, maynard_82}. We note that this is not always the case when modeling human interactions, who can change behaviors depending also on other external, and to a large degree unpredictable, factors. We also note that many works have shown that the strategy updating rule can have profound influence on the evolution of strategies, even changing the impact of the topology of interaction network \cite{szabo_pr07,perc_bs10, gracia-lazaro_pnas12, hauert_n04}.

\subsection{Triangular lattice}

\begin{figure}
\centerline{\epsfig{file=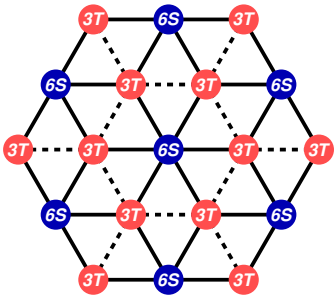,width=5cm}}
\caption{(Color online) The minimal frustration configuration in the triangular lattice. Every blue (dark gray) site (cooperator) receives the best payoff, although defectors, marked red (light gray), have only half of their connections leading to the best payoff. Frustrated bonds are drawn by dashed lines and bonds that maximize the payoff are drawn by full lines.}
\label{fig_pattern}
\end{figure}

We make a quick review here to clarify some properties of the triangular lattice. This topology has an important property: every closed loop is comprised of an even number of steps, which gives rise to frustration phenomena \cite{Robinson2011}. The snowdrift game, which is also known as anti-coordination game since choosing the opposite strategy of the partner  is a Nash-equilibrium, is strongly affected by network inherent frustration.  In square lattices, the logit rule yields a population displaying a very stable checkerboard pattern, as everyone can choose to do the opposite of all neighbors \cite{szabo_pr07, sysiaho_epjb05, hauert_n04, Choi2015, weisbuch_pa07, szabo_jtb12, szabo_jtb12b, blume_l_geb93}. In contrast, this spatial ordering is impossible in the triangular lattice, as shown in Fig.~\ref{fig_pattern}. Every pair of different strategies will share at least one third neighbor that will be frustrated. This phenomenon is well explored in magnetic models, where many interesting ``spin-glass'' phenomena can arise \cite{nishimori_01, Robinson2011}. In spin models we see that the ``minimum energy'' configuration would be similar to the pattern shown in Fig.~\ref{fig_pattern}. We wish to analyze this situation in evolutionary game dynamics. One type of player is surrounded by a honeycomb structure of the opposite type, repeated infinitely for a large lattice. Notice that the central site (blue) does not have any frustrated connections, while the other type (red) is frustrated in half of its connections.

\section{Results}

We start showing that, mathematically, the  formal relation between anti-coordination games and anti-ferromagnetic systems \cite{szabo_jtb12b, szabo_jtb12, szabo_pr07, szabo_pre05, blume_l_geb93, szabo_pre10, weisbuch_pa07,  galam_pa10, hauert_n04, sysiaho_epjb05} is not an identity. Let us consider a matrix for the energy of a single spin in a magnet with coupling constant $J$ and an external magnetic field $B$, similar to the payoff matrix~\ref{paymatrix}:
\begin{equation}
 \bordermatrix{~ & \uparrow & \downarrow \cr
                  \uparrow & -J-B & J-B \cr
                  \downarrow & J+B & -J+B \cr}.
\end{equation}
 The spins in the anti-ferromagnet ($J<0$) tend to point in the opposite direction as their neighbors, as in anti-coordination games individuals tend to do the opposite of their neighbors. However, equating the payoff matrix to the energy matrix ($-J-B=R$, $J-B=S$, $J+B=T$, and $-J+B=P$) and requiring the snowdrift payoff condition  ($T>R>S>P$) yield
\begin{equation}
J+B>0 , \quad 0>J \quad \textrm{and}  \quad J>B ,
\end{equation}
which is a mathematical absurd. There is no combination of parameters that obey both the physical symmetry of magnetic system and the dilemma hierarchy of game theory for a general case. In other words, the magnetic system obeys a diagonal symmetry in the matrix, whereas the game theory obeys a linear hierarchy of the parameters in the matrix, both cannot be  fulfilled simultaneously. It is important to stress that, although we will see many phenomena in the simulations that are analogous to anti-ferromagnetism, the systems are not formally identical.

\subsection{Master equation}

Let us analyze the logit model using mean-field approximation at nearest-neighbor level \cite{szabo_pr07, matsuda_h_ptp92, schuster_jtb83}. For simplicity we set $S=0$ in this section. If $T>1$, we have the so-called weak prisoner's dilemma. Consider a central site $i$ on a lattice. It interacts only with its four (square lattice) or six (triangular lattice) nearest neighbors ($\Omega$ neighborhood). In this setup, we present the master equation  for the average fraction of cooperators, $\rho$ (note that $\rho$ is a function of $t$):
\begin{equation}\label{master}
\dot{\rho}= (1-\rho) \Gamma_{+(C\rightarrow D)} - \rho \Gamma _{-(D\rightarrow C)}
\end{equation}
where $\Gamma{\pm}$ is the probability for the central player to change its strategy to $C$ ($D$).
We obtain ${{N}\choose{n}}$ different neighborhood configurations where $N$ is $4$ for the square lattice and $6$ for the triangular lattice and $n$ is the number of cooperative neighbors for each neighborhood configuration. Therefore:
\begin{equation}\label{gamma}
\Gamma_{\pm}= \sum_{n=0}^{N} {{N}\choose{n}} \rho^n (1-\rho)^{(N-n)} P_{\pm}(u_i,u_{\Omega}).
\end{equation}

Here, ${{N}\choose{n}}$ are the binomial coefficients and weights the repetitions of identical configurations. Note that while $n$ varies in the summation, $N$ is fixed for each lattice type. The term $\rho^n (1-\rho)^{(N-n)}$ weights the probability of such configuration and $P_{\pm}(u_i,u_{\Omega})$ is the probability, in a specific configuration, that the central site will turn into a cooperator ($P_{+}$) or a defector ($P_{-}$). This probability is the only term that is directly dependent on the update rule chosen (logit or imitation). For the logit rule the focal site changes the state comparing its current payoff ($u_i$) with its future payoff if the state was changed, ($u^*$). Calculating $P_{+}(u_i,u_{\Omega})$, for the case where the central site is $D$ and changes to $C$, we have:
\begin{equation}
P_{+}(u_i,u_{\Omega})=\frac{1}{1+e^{-(u^*-u_i)/K}}.
\end{equation}

Analytically, one of the advantages of the logit model is that the probability does not depend explicitly on the payoffs of the neighborhood $\Omega$. If the central site is $D$ ($C$), the payoff difference, for any configuration,  will be:
\begin{eqnarray}
(u^*-u_i)_{D \rightarrow C}=n(1-T) ~, \\
(u^*-u_i)_{C \rightarrow D}=n(T-1).
\end{eqnarray}
Using $A=(1-T)/K$ to simplify, we get:
\begin{equation}
P_{\pm}(u_i,u_{\Omega})=\frac{1}{1+e^{\mp n A}}.
\end{equation}

Remember that the solution for the master equation of the imitation model can be found in the literature \cite{szabo_pr07, nowak_06, amaral_jpa15}. The master equation for the logit model becomes:
\begin{equation}\label{ising_master}
\dot{\rho}=\sum_{n=0}^{N}{{N}\choose{n}} \rho^n (1-\rho)^{(N-n)}\left(\frac{1}{1+e^{\frac{- n (1-T)}{K}}} -\rho\right).
\end{equation}

This yields a 6th order polynomial that analytically have at least one root in the region $0<\rho ^*<1$. This is independent of $T$, meaning that at the nearest-neighbor level there exist at least some minimum cooperation level independently of the value of temptation. The existence of a minimum level of cooperation is an interesting result, agreeing with other approaches on innovative dynamics that found similar results using Monte Carlo simulations and experiments with humans \cite{roca_epjb09, szolnoki_pre14, fort_jsm05, szabo_jtb12b, szabo_jtb12, szabo_pre10, gracia-lazaro_pnas12, gracia-lazaro_srep12, Grujic2014}.

\begin{figure}
\centerline{\epsfig{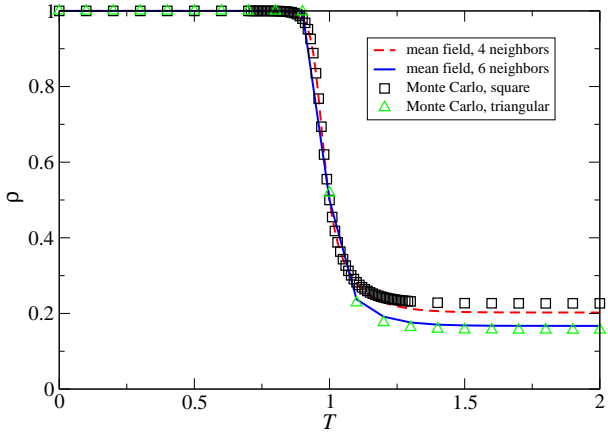}}
\caption{(Color online) The fraction of cooperators in equilibrium as a function of $T$. The results refer to Monte Carlo simulations (symbols) and the master equation ODE (lines) in the logit model for both lattices. Note that simulation and analytical results agree well and reproduce the main characteristic of the system, namely a non-vanishing cooperation level.}
\label{fig_isingnum}
\end{figure}

To obtain the time-independent solution of the master equation we use a 4th order Runge-Kutta integrator. As in other models, the system reaches a stable state after some time. In our model the behavior of $\rho(t)_{t\rightarrow \infty}$ is independent of the initial fraction of cooperation. This is an important feature, as not every update rule will have a equilibrium state independent of the initial conditions \cite{roca_epjb09, wang_z_srep12, vainstein_pre01, arapaki_pa09}. Figure \ref{fig_isingnum} shows the cooperation level for the stable equilibrium ($\rho(t)_{t \rightarrow \infty}$) as a function of $T$. We compare the Monte Carlo simulation (further analyzed bellow) with the numerical solution for the master equation in both topologies. The mean field approach agrees with the simulation results and, most importantly, both approaches report a basal cooperation level for any $T$.

The mean-field technique is a good approximation to obtain insights and confirm the prediction of other methods. Even so, it does not always returns the same results as in the structured population \cite{szabo_pr07, szabo_pre05}, it is only an approximation. In our case, it is interesting to notice that both methodologies (Monte Carlo and mean field) report the minimal level of cooperation that is independent of the value of temptation. This kind of basal cooperation level was also found in other studies using innovative dynamics, even with different update rules and topologies \cite{szabo_jtb12b, szabo_jtb12, szabo_pre10}.

\subsection{Monte Carlo simulations}

We use the asynchronous Monte Carlo procedure to simulate the evolutionary dynamics. First, a randomly chosen player, $i$, is selected. The cumulative payoff of $i$ and of its nearest neighbors payoffs are calculated. Then player $i$ changes its strategy based on the update probability defined in Eq.~\ref{isingeq} for logit or in Eq.~\ref{imiteq} for imitation dynamics. One Monte Carlo step ($MCS$) consists of this process repeated $L^2$ times, where $L$ is the lattice linear size (here we set $L=100$). For a detailed discussion on Monte Carlo methods in evolutionary dynamics we suggest Refs.~\cite{binder_88, binder_rpp97, szabo_pr07, huberman_pnas93}. We ran the algorithm until the equilibrium state ($10^4-10^5$ $MCS$'s); then we average over 1000 $MCS$'s for $10-20$ different initial conditions. We used periodic boundary conditions and random, homogeneous initial strategy distribution.

\begin{figure}
\centerline{\epsfig{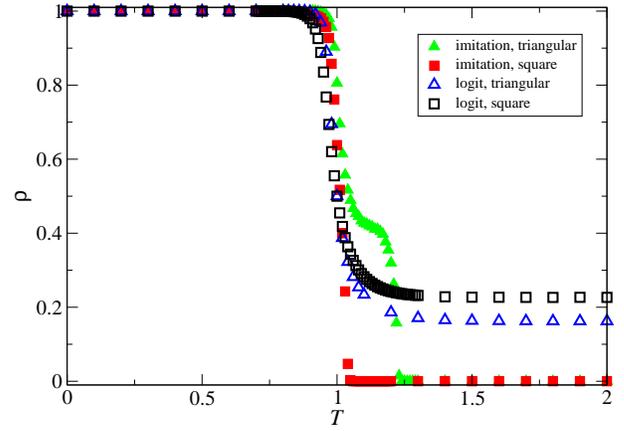}}
\caption{(Color online) The fraction of cooperators in equilibrium as a function of $T$, as obtained by means of Monte Carlo simulations (for $S=0$). Note that the difference between the logit and the imitation model is significant for $T>1$, where the logit model exhibits a minimal cooperation level. Also note that the sharp drop in cooperation occurs in the same region.}
\label{fig_comp2D}
\end{figure}

Starting with the weak prisoner's dilemma ($S=0$), we compare the logit with the imitative dynamics. Figure~\ref{fig_comp2D} shows  $\rho$ as a function of $T$. The logit model has a sharp decay in cooperation, almost at the same point where the imitation model has a transition \cite{szabo_pre05,szabo_pr07}. This is valid for both square and triangular lattices. Also, it is remarkable that for large $T$ a minimal global value of cooperation survives, confirming the prediction of our mean field approach.

\begin{figure}
\centerline{\epsfig{file=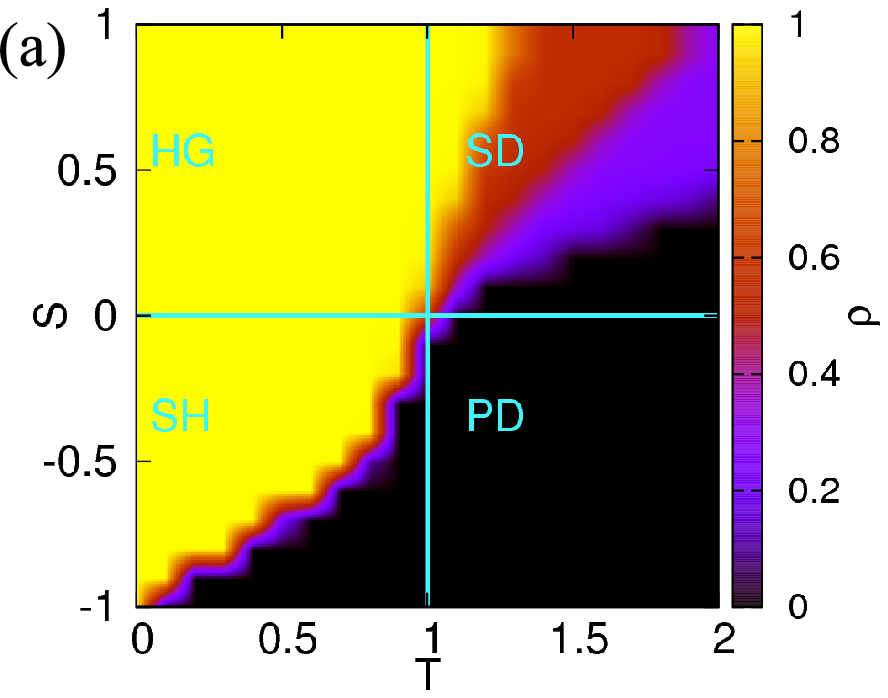,width=4.5cm}\epsfig{file=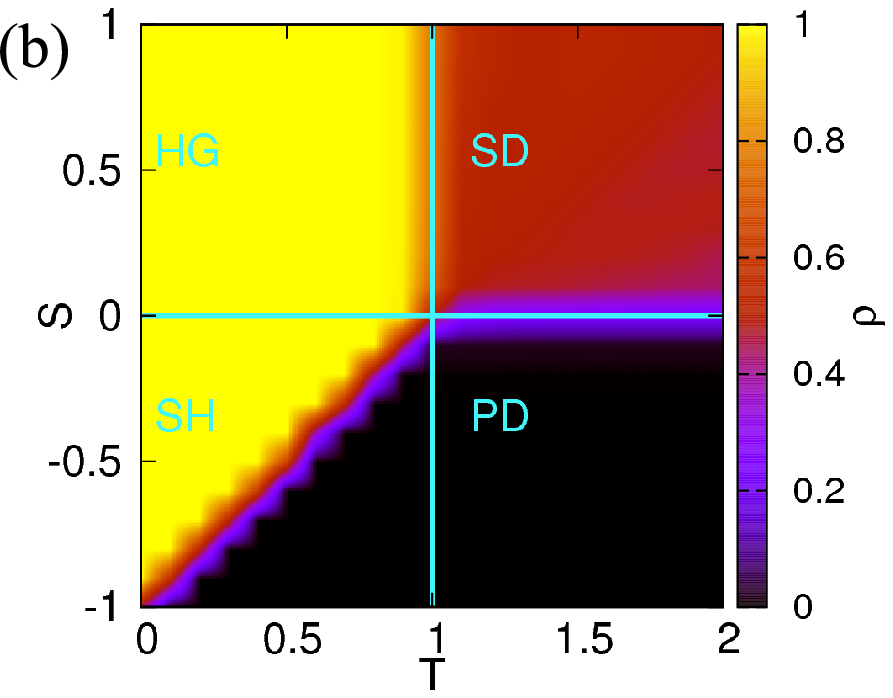,width=4.5cm}}
\centerline{\epsfig{file=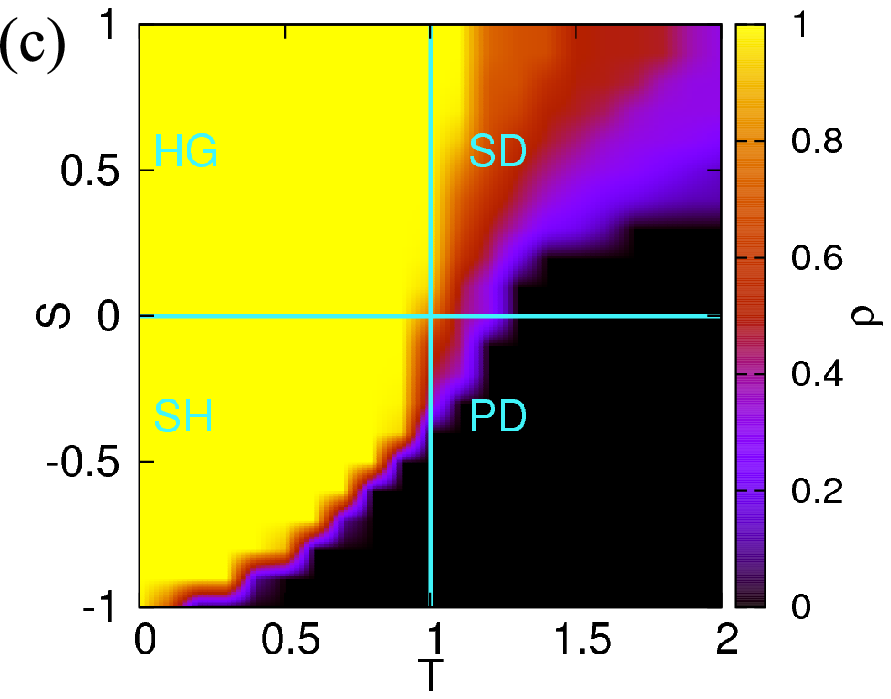,width=4.5cm}\epsfig{file=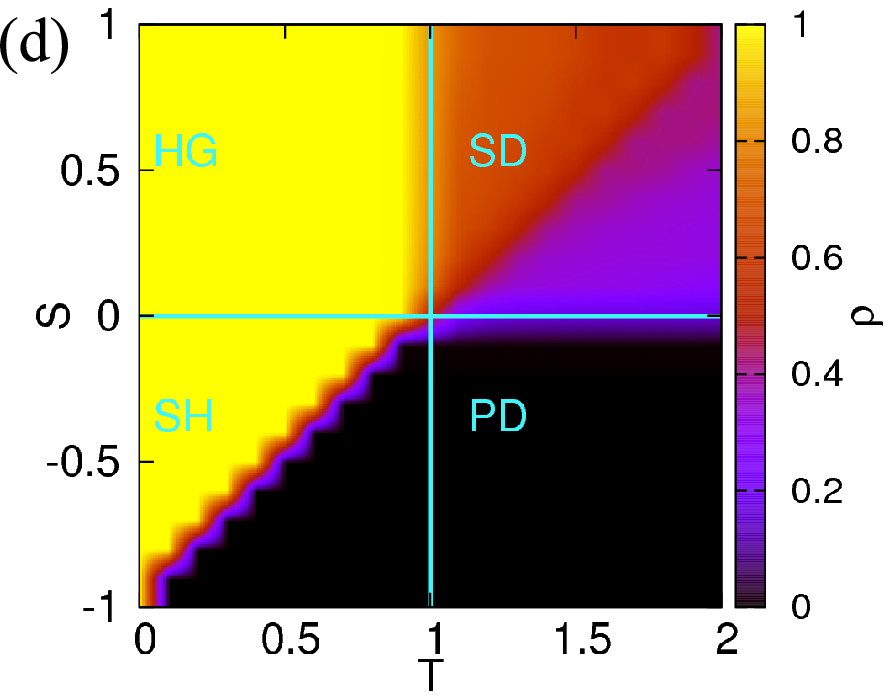,width=4.5cm}}
\caption{(Color online) Heat maps encoding the cooperation level for the whole $T-S$ plane. The top row, (a) and (b), shows results obtained on the square lattice, while the bottom row, (c) and (d), shows results obtained on the triangular lattice. The left column, (a) and (c), shows results obtained with the imitation dynamics, and the right column, (b) and (d), shows the results obtained with the logit rule. When imitation dynamics is used, there is little difference inferable that would be due to the differences in the interaction lattice. For the logit model, the level of cooperation is higher in the SD region for both topologies. Most interestingly, for the triangular lattice, we can observe two different phases that are separated by a straight line.}
\label{fig_2x2phases}
\end{figure}

Figure~\ref{fig_2x2phases} shows the fraction of cooperation in the entire $T-S$ plane in the imitation and logit models, for both triangular and square lattices. Notice how similar the outcomes are in the HG, PD and SH games. The  difference appears  in the snowdrift game. Imitation dynamics yields similar results in both square and triangular lattices, but  logit dynamics yields different results. More specifically, while in the logit model on the square lattice there is a flat plateau of $50\%$ cooperation (deeply studied in \cite{szabo_jtb12b, szabo_jtb12, szabo_pre10}), in the triangular lattice there are basically two phases separated by a straight diagonal line ($S=T-1$). Notice that on the square lattice the whole SD region is associated with a static checkerboard pattern, corresponding to the Nash Equilibrium, which is the  most efficient way of increasing the population payoff, as previously stressed in \cite{szabo_jtb12b, szabo_jtb12, szabo_pre10, fort_jsm05, weisbuch_pa07, roca_pre09, sysiaho_epjb05}). It is also interesting to notice that for the logit model, cooperation survives independently of $T$ for some range of $S$ (around $S\simeq-0.15$) in the PD region.

\begin{figure}
\centerline{\epsfig{file=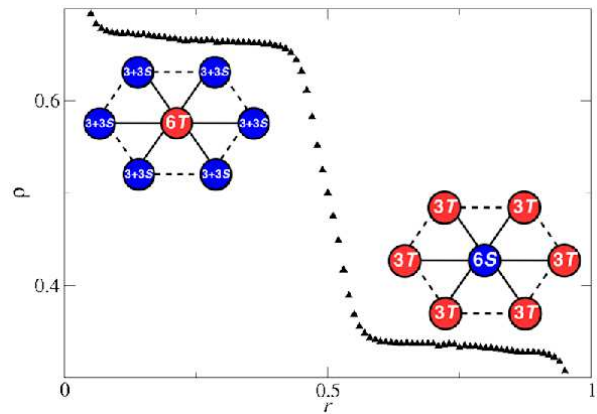,width=7.9cm}}
\caption{(Color online) The fraction of cooperators in equilibrium along the line defined by $T=2-S$ in the logit model on a triangular lattice. Here the payoff values are varied via the control parameter $r$, where $T=1+r$ and $S=1-r$. Instead of a homogeneous state, like on the square lattice, we observe two different phases with honeycomb-like spatial patterns. The insets illustrate the typical honeycomb cell that is characteristic of each phase.}
\label{fig_diagonal}
\end{figure}

Studying the SD region for triangular lattice in the logit dynamics, we find a plateau of $\rho \simeq 0.35$ bellow the diagonal line and $\rho \simeq 0.65$ above it, with minor fluctuations of $\pm 0.05$. We further refer to Refs.~\cite{szabo_jtb12b, szabo_jtb12, szabo_pre10, weisbuch_pa07} for the analysis of  the square lattice, where such plateau is also found with a single phase ($\rho=0.5$).  In principle, there would be two ``ground states'' exhibiting a honeycomb pattern: a concatenation of cells with a central $D$ surrounded by $C$'s and a concatenation of cells with a central $C$ surrounded by $D$'s. Let us consider the first ``ground state'', where the central site in each cell of the honeycomb configuration is a defector surrounded by 6 cooperators. Each one of these 6 cooperators is shared by 3 distinct cells. The fraction $\rho$ in an infinite lattice is calculated as the fraction of cooperators in the cell, weighting each site by the number of blocks which share it. So we have
\begin{equation}
 \rho=\frac{6/3}{6/3+1}=\frac{2}{3}
\end{equation}
The calculation for the other ``ground state'' is  analogous, yielding $\rho=1/3$. Most interestingly, the systems is driven to one of the two ``ground states'' configurations  depending on the payoff parameters. To make this point clearer, in Fig.~\ref{fig_diagonal} we show the fraction of cooperation for parameters along a straight line orthogonal to the line that divides the plateaus observed in the SD region. We can clearly see the two plateaus and the transition point where the roles of $C$ and $D$ players are exchanged, as shown in the incept of the patterns.

\begin{figure}
\centerline{\epsfig{file=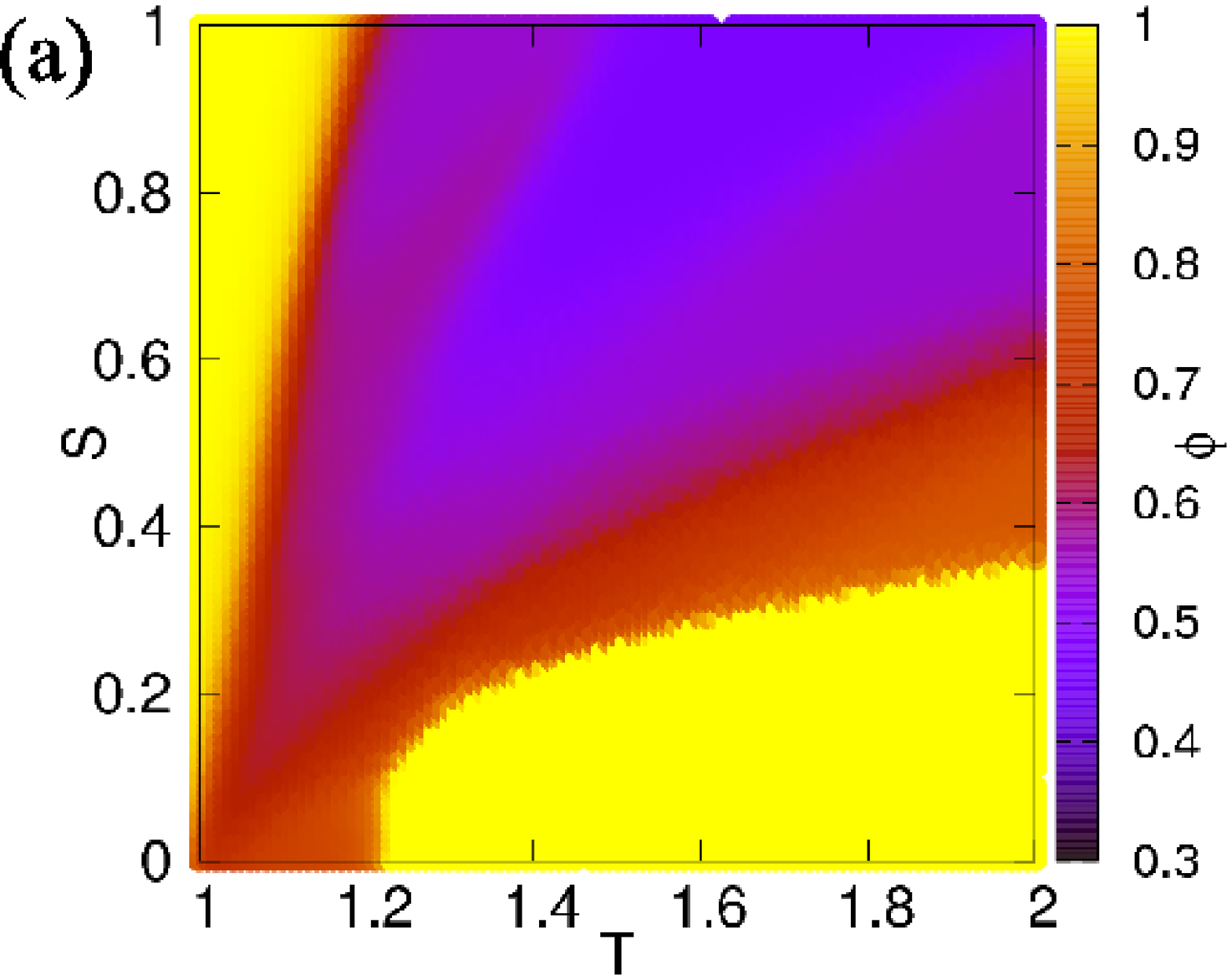,width=4.5cm}\epsfig{file=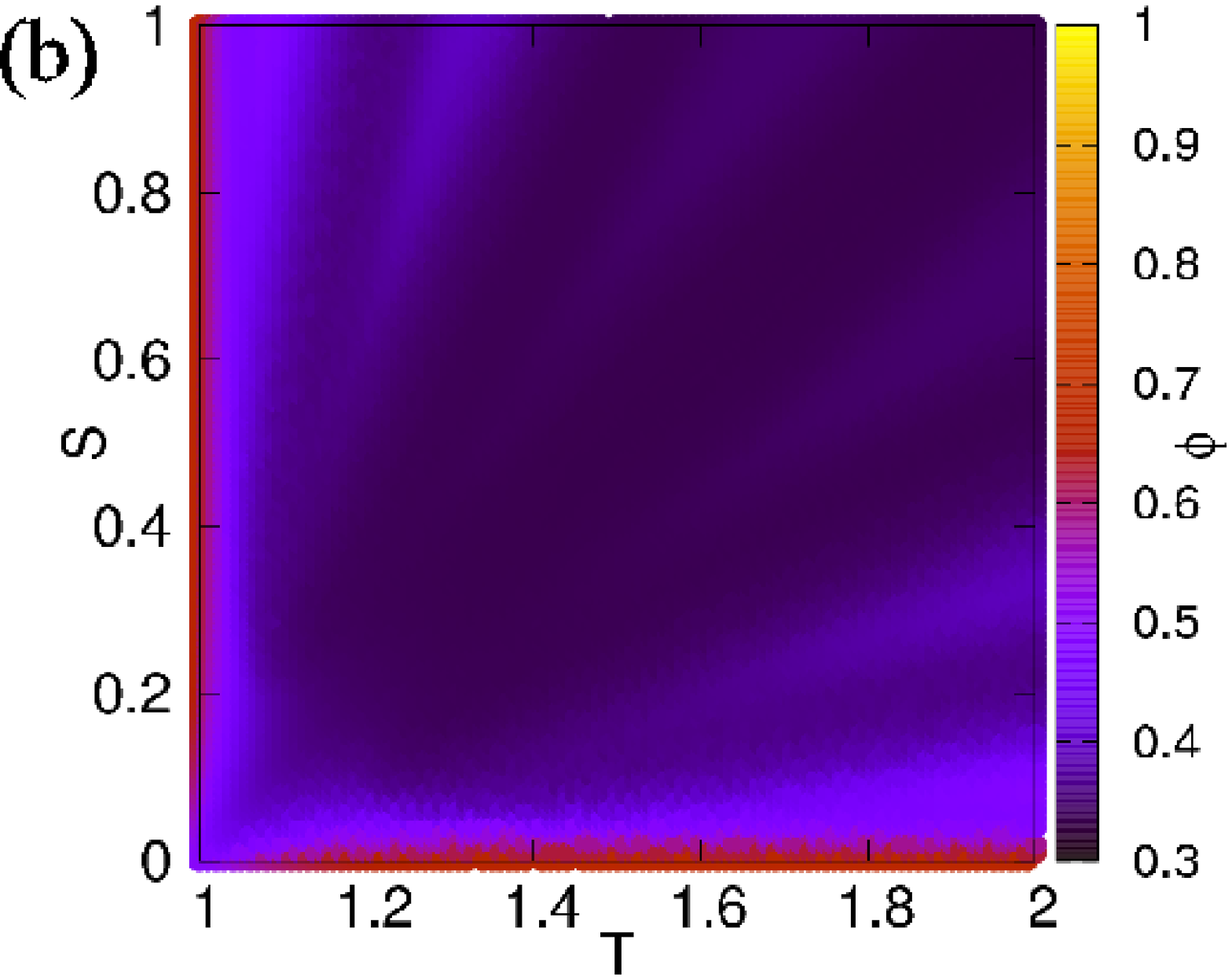,width=4.5cm}}
\caption{(Color online) The level of frustration, $\phi$,  in the snowdrift region of the $T-S$ diagram for the imitation (a) and logit (b) strategy updating rules on the triangular lattice. The imitation model has many frustrated links, around $60\%$, while the logit model maintains the low and homogeneous frustration of around $35\%$.}
\label{fig_frustphase}
\end{figure}

The logit model seems to drive the system to the maximum attainable global payoff (related to the minimum energy level). To further study this hypothesis, we  quantify the frustration, $\phi$, defined as the fraction of frustrated links. In SD games the frustrated links are the $CC$ and $DD$ pairs. Note that our definition of frustration  is a good measurement of the ``homogeneity'' and global spatial structure of the lattice: the frustration is $1$ for any homogeneous state, regardless of the cooperation level, and can be zero, for example, in the chess board pattern configuration of cooperators and defectors on square lattices. In both ``ground states'' configuration of the triangular lattice, we can easily show that the frustration is equal to $1/3$. In Fig.~\ref{fig_frustphase} we compare the lattice frustration of logit and imitation rules for the SD region (frustration is meaningless outside this parameter range). The imitation model maintains a high frustration, around $60\%$, whereas  the logit model maintains a moderate frustration, around $35\%$, independently of payoff values of $T$ or $S$, which is very close to the  analytical solution of the honeycomb structure. Note that on triangular lattices the minimum achievable frustration is $1/3$, as there is an inevitable  topological frustration. Also note how  frustration quickly rises to almost $1$ in the borders of the diagram, where there is full cooperation or full defection.

\begin{figure}
\centerline{\epsfig{file=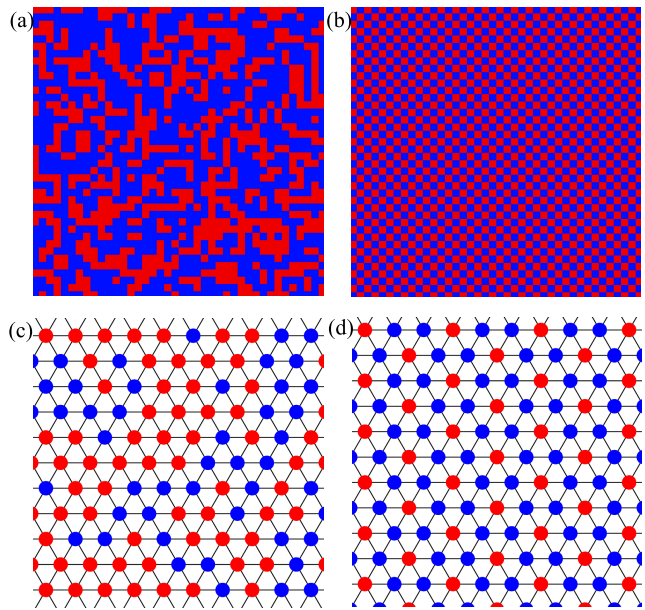,width=8.5cm}}
\caption{(Color online) Typical snapshots of the square lattice in the top row, (a) and (b), and the triangular lattice in the bottom row, (c) and (d), in the SD region. The left column, (a) and (c), shows the results for imitation model while the right column, (b) and (d), shows the results for the logit model. In the logit model on the square lattice a checkerboard pattern quickly emerges. In the logit model on the triangular lattice, on the other hand, we see the honeycomb pattern. Here we use $T=1.2$ and $S=0.5$.}
\label{fig_2x2snapshot}
\end{figure}

To further support our claims, we present snapshots of the lattices in dynamic equilibrium on the SD region. The Monte Carlo method is of course probabilistic, and accurate results are dependent on sufficiently large averages \cite{binder_88, binder_rpp97, szabo_pr07, huberman_pnas93}. Even so, it is insightful to see the images of the lattice after the system has reached a dynamical equilibrium. Figure~\ref{fig_2x2snapshot} shows typical snapshots of logit and imitation update rules for square and triangular lattices. It is clear the differences in spatial organization exhibited in each model. We see that in both topologies the imitation update tends to maintain cooperators in clusters whereas the logit model tends to distribute strategies more homogeneously. Specifically, the logit model on square lattice tends to form a checkerboard pattern, a behavior that has been consistently reported in different innovative rules \cite{fort_jsm05, szabo_jtb12b, szabo_jtb12, szabo_pre10, weisbuch_pa07, roca_pre09, sysiaho_epjb05} and is usually attributed to the population re-arranging itself to receive the highest total payoff achievable. For the triangular lattice we can see that the expected frustrated pattern illustrated in Fig.~\ref{fig_pattern} indeed emerges. It is worth mentioning that  it is the absence of clustering that makes the mean-field approximation a good one for the logit dynamics. Such phenomena suggests a general behavior exhibited by innovative dynamics that leads to the emergence of specific spatial structures, other than cooperation islands. We note that, while clustering has a strong effect on everyday cooperative interactions \cite{Buonanno2009, Botzen2016}, the emergence of diluted patterns in our model suggests that some role-separating structure may also emerge in human population, where members have different roles to obtain a higher collective income.

\begin{figure}
\centerline{\epsfig{file=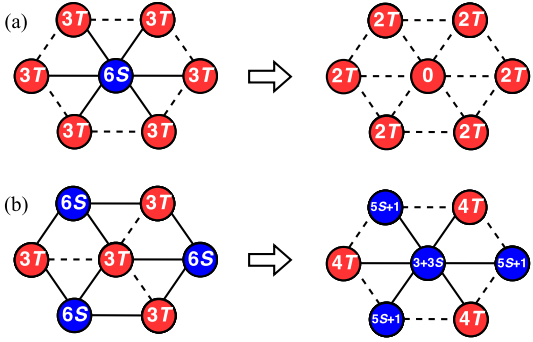,width=7cm}}
\caption{(Color online) Leading microscopic processes that guide pattern formation in a frustrated topology. (a) A cooperator surrounded by defectors is very stable, since the change in payoff here would be $-6S$. (b) Defectors in the vertices, now shown in the center, may change their strategy, depending on the parameters. The payoff difference would be $3(S-T+1)$. If $S>T-1$ the chance that defector becomes a cooperator is high. $S=T-1$ is the line dividing the two phases seen in Fig.~\ref{fig_2x2phases}.}
\label{fig_trimicro}
\end{figure}

Lastly, we analyze the representative microscopic mechanisms that explain how strategy evolution accommodates to topological frustration. In Fig.~\ref{fig_trimicro}(a) we present a local strategy distribution where the central site is highly unlikely to change its strategy that makes the honeycomb configuration very stable. Conceptually similar stable local distribution can be drawn where a defector is surrounded by cooperators. However, the sites around the central stable site are not fully satisfied because they have some frustrated bonds. This situation is illustrated in Fig.~\ref{fig_trimicro}(b) where a frustrated node is in the center. Here the central site has a higher chance to change its strategy depending on the difference between $(3T)$ and $(3+3S)$. The threshold value is at the line $S=T-1$ which agrees perfectly with the border line we observed in Fig.~\ref{fig_2x2phases}. The frustrated sites have a pivotal role in the separation of phases illustrated in Fig.~\ref{fig_diagonal}. For low $T$ values, that is for $S>T-1$, cooperators fare better than defectors, allowing them to stay in ``frustrated'' sites of the honeycomb configuration. This results in a large number of cooperators, as every defector will be surrounded by 6 cooperators ($\rho \sim 2/3$ in the infinitely repeated limit). The opposite is also true for $S<T-1$, namely, the defectors have a high payoff, allowing them to stay in the frustrated positions of the honeycomb patches. As a result, a stable cooperator will be surrounded by 6 defectors yielding a relatively low cooperation level ($\rho \sim 1/3$ in the infinite limit).

We found that frustration can induce two distinct organized patterns on triangular lattices. As we noted, square lattice topology can be considered as the opposite extreme case where there is no frustrated bonds between players. We wonder how these extreme cases can be bridged by an appropriately modified topology where the frustration level can be tuned gradually. To generate such an intermediate level of inherent frustration we modify the triangular lattice by removing two diagonal connections of each site. When we alter the originally triangular lattice then the control parameter is the $X$ fraction of sites that have their diagonal links removed. Accordingly, $X=0$ corresponds to the triangular lattice while at $X=1$ the resulting topology agrees with the square lattice. Note that  the network remains static throughout the evolutionary process and we study how the strategy evolution may change due to the intermediate level of topological frustration.

\begin{figure}
\centerline{\epsfig{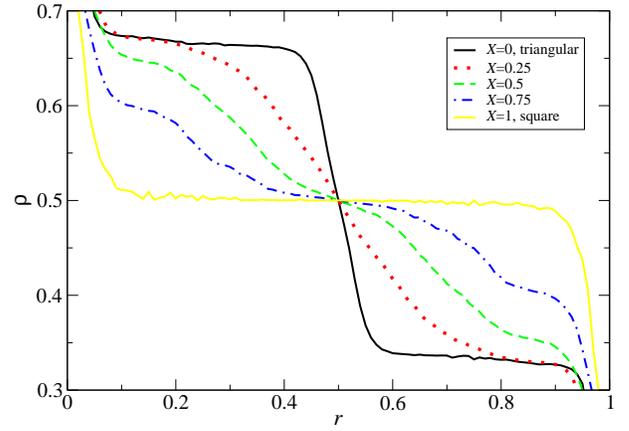}}
\caption{(Color online) The effect of link deletion on the cooperation level along the $T=2-S$ line in the logit model. Here the payoff values are varied via the control parameter $r$, where $T=1+r$, while $S=1-r$. The topology is modified gradually where a fraction $X$ of diagonal links are removed from a triangular lattice. Accordingly, $X=0$ corresponds to the triangular lattice and $X=1$ to the square lattice topology. If the topological frustration is mitigated by deleting just a few links, then the steep transition between ordered phases vanishes. Alternatively, when we add a tiny frustration to the interaction graph by leaving $X=1$ then the anti-ferromagnetic order disappears immediately.}
\label{fig_topology}
\end{figure}

Figure~\ref{fig_topology} shows the resulting cooperation level in dependence on the payoff values for differently frustrated topologies as characterized by the value of $X$. As we start mitigating the maximal frustration by increasing $X$, the steep transition point separating the two plateaus vanishes immediately verifying that the two ordered phases can only be observed when maximal level of frustration is present in the topology. As we increase $X$ further then the resulting $\rho(r)$ function will approximate the $\rho=0.5$ plateau only at the $X \to 1$ limit. It simply means that the long-range anti-ferromagnetic order of competing strategies disappears immediately when we leave the $X=1$ point and introduce some frustration into the perfectly frustration-free square lattice topology. In between these extreme cases the shape of the $\rho(t)$ function may inform us about the frustration level of the unknown interaction graph.

\section{Discussion}
In social interactions the ``best response'' is often challenging, especially if the interaction involves a triangle. In general, frustrated situations can arise as a consequence of the type of game played, due to specific interaction topologies, but also because of other external factors. Motivated by this phenomenon, we considered the snowdrift game on a triangular lattice where the topological frustration inhibits the expected optimal anti-coordination of strategies. By means of master-equation approximations and Monte Carlo simulations, we have studied the logit strategy updating protocol, the classical imitation dynamics, and we have compared the evolutionary outcomes obtained on the triangular lattice, the square lattice, and on an abridged transition between the two that was achieved by randomly adding links to the next-nearest neighbors of the square lattice. Our principal interest was to reveal how topological frustration influence the strategy ordering in a spatial system.

In stark contrast to the square lattice where anti-coordination ordering can emerge, the frustrated topology of triangular lattice generates two ordered phases in the snowdrift quadrant. These states are separated by the $S=T-1$ line. While for low $T$ values cooperators occupy $2/3$ of the available sites and the rest is occupied by defectors, their roles are exchanged for high $T$ values. In both phases the system evolves into a state which is reminiscent to a honeycomb-like pattern that helps to minimize the negative consequence of the topological frustration. We have identified the microscopic mechanisms which compose these patterns, and we have found that such formations are very stable. By comparing them with the outcome of imitation dynamics, we have found that the logit rule allows the whole system to evolve into the least frustrated strategy distribution that is achievable on each lattice, which also provides the highest population payoff. This state is reached via a strategy distribution where cooperators are less clustered comparing to the patterns constructed by imitation dynamics.

The striking difference between frustration-free (square) and frustrated (triangle) lattices raises a question on what we shall expect if the interaction graph is disordered and the level of topological frustration is unknown. What kind of behavior is expected in such a case? To clarify this, we have introduced a method which allowed us to modify the level of topological frustration gradually. Starting from a triangular lattice, we have randomly deleted a fraction of diagonal links, which decreased the frustration between neighboring bonds. If all diagonal links were deleted, then we arrive at the square lattice. We found that the two ordered phases in the snowdrift quadrant disappear as we mitigate the frustration level. On the other hand, the well-known anti-ferromagnetic like checkerboard pattern observed on a square lattice, which is valid for the whole scan of the mentioned quadrant of the $T-S$ plane, evaporates immediately as we introduce a tiny frustration into the interaction topology. These phenomena highlight how frustration can drive individuals to form complex global patterns, and more importantly, how innovative dynamics can drive the system to the best, i.e., least frustrated, evolutionary outcome.

As the strategy updating rule can drastically alter population dynamics, it is important to study how different protocols deal with frustration and which kind of patterns can spontaneously emerge from the applied dynamic. This is even more interesting in the light of emergence of complexity as individuals interact. The studied logit rule model is essential to the emergence of the patterns shown here, and recent research shows the importance of integrating innovative dynamics in game theoretical models, especially since humans seem to use different rules than simply imitating the best when playing evolutionary games \cite{gracia-lazaro_pnas12, gracia-lazaro_srep12, Grujic2014, rand_nc14, capraro2014heuristics, capraro2015social, bear_pnas16, roca_epjb09}. We hope that this paper will motivate further research along this area in the future.

\begin{acknowledgments}
This research was supported by the Brazilian Research Agencies CAPES-PDSE (Proc. BEX 7304/15-3), CNPq and FAPEMIG, by the Slovenian Research Agency (Grants J1-7009 and P5-0027), and by the Hungarian National Research Fund (Grant K-120785).
\end{acknowledgments}


\begin{thebibliography}{111}
\expandafter\ifx\csname natexlab\endcsname\relax\def\natexlab#1{#1}\fi
\expandafter\ifx\csname bibnamefont\endcsname\relax
  \def\bibnamefont#1{#1}\fi
\expandafter\ifx\csname bibfnamefont\endcsname\relax
  \def\bibfnamefont#1{#1}\fi
\expandafter\ifx\csname citenamefont\endcsname\relax
  \def\citenamefont#1{#1}\fi
\expandafter\ifx\csname url\endcsname\relax
  \def\url#1{\texttt{#1}}\fi
\expandafter\ifx\csname urlprefix\endcsname\relax\def\urlprefix{URL }\fi
\providecommand{\bibinfo}[2]{#2}
\providecommand{\eprint}[2][]{\url{#2}}

\bibitem[{\citenamefont{Pennisi}(2005)}]{pennisi_s05}
\bibinfo{author}{\bibfnamefont{E.}~\bibnamefont{Pennisi}},
  \bibinfo{journal}{Science} \textbf{\bibinfo{volume}{309}},
  \bibinfo{pages}{93} (\bibinfo{year}{2005}).

\bibitem[{\citenamefont{Maynard~Smith}(1982)}]{maynard_82}
\bibinfo{author}{\bibfnamefont{J.}~\bibnamefont{Maynard~Smith}},
  \emph{\bibinfo{title}{Evolution and the Theory of Games}}
  (\bibinfo{publisher}{Cambridge University Press},
  \bibinfo{address}{Cambridge, U.K.}, \bibinfo{year}{1982}).

\bibitem[{\citenamefont{Weibull}(1995)}]{weibull_95}
\bibinfo{author}{\bibfnamefont{J.~W.} \bibnamefont{Weibull}},
  \emph{\bibinfo{title}{Evolutionary Game Theory}} (\bibinfo{publisher}{MIT
  Press}, \bibinfo{address}{Cambridge, MA}, \bibinfo{year}{1995}).

\bibitem[{\citenamefont{Hofbauer and Sigmund}(1998)}]{hofbauer_98}
\bibinfo{author}{\bibfnamefont{J.}~\bibnamefont{Hofbauer}} \bibnamefont{and}
  \bibinfo{author}{\bibfnamefont{K.}~\bibnamefont{Sigmund}},
  \emph{\bibinfo{title}{Evolutionary Games and Population Dynamics}}
  (\bibinfo{publisher}{Cambridge University Press},
  \bibinfo{address}{Cambridge, U.K.}, \bibinfo{year}{1998}).

\bibitem[{\citenamefont{Mesterton-Gibbons}(2001)}]{mestertong_01}
\bibinfo{author}{\bibfnamefont{M.}~\bibnamefont{Mesterton-Gibbons}},
  \emph{\bibinfo{title}{An Introduction to Game-Theoretic Modelling, 2nd
  Edition}} (\bibinfo{publisher}{American Mathematical Society},
  \bibinfo{address}{Providence, RI}, \bibinfo{year}{2001}).

\bibitem[{\citenamefont{Nowak}(2006)}]{nowak_06}
\bibinfo{author}{\bibfnamefont{M.~A.} \bibnamefont{Nowak}},
  \emph{\bibinfo{title}{Evolutionary Dynamics}} (\bibinfo{publisher}{Harvard
  University Press}, \bibinfo{address}{Cambridge, MA}, \bibinfo{year}{2006}).

\bibitem[{\citenamefont{Axelrod}(1984)}]{axelrod_84}
\bibinfo{author}{\bibfnamefont{R.}~\bibnamefont{Axelrod}},
  \emph{\bibinfo{title}{The Evolution of Cooperation}}
  (\bibinfo{publisher}{Basic Books}, \bibinfo{address}{New York},
  \bibinfo{year}{1984}).

\bibitem[{\citenamefont{Wilson}(1971)}]{wilson_71}
\bibinfo{author}{\bibfnamefont{E.~O.} \bibnamefont{Wilson}},
  \emph{\bibinfo{title}{The Insect Societies}} (\bibinfo{publisher}{Harvard
  Univ. Press}, \bibinfo{address}{Harvard}, \bibinfo{year}{1971}).

\bibitem[{\citenamefont{Skutch}(1961)}]{skutch_co61}
\bibinfo{author}{\bibfnamefont{A.~F.} \bibnamefont{Skutch}},
  \bibinfo{journal}{Condor} \textbf{\bibinfo{volume}{63}}, \bibinfo{pages}{198}
  (\bibinfo{year}{1961}).

\bibitem[{\citenamefont{Nowak and Highfield}(2011)}]{nowak_11}
\bibinfo{author}{\bibfnamefont{M.~A.} \bibnamefont{Nowak}} \bibnamefont{and}
  \bibinfo{author}{\bibfnamefont{R.}~\bibnamefont{Highfield}},
  \emph{\bibinfo{title}{SuperCooperators: Altruism, Evolution, and Why We Need
  Each Other to Succeed}} (\bibinfo{publisher}{Free Press},
  \bibinfo{address}{New York}, \bibinfo{year}{2011}).

\bibitem[{\citenamefont{Nowak and May}(1992)}]{nowak_n92b}
\bibinfo{author}{\bibfnamefont{M.~A.} \bibnamefont{Nowak}} \bibnamefont{and}
  \bibinfo{author}{\bibfnamefont{R.~M.} \bibnamefont{May}},
  \bibinfo{journal}{Nature} \textbf{\bibinfo{volume}{359}},
  \bibinfo{pages}{826} (\bibinfo{year}{1992}).

\bibitem[{\citenamefont{Hamilton}(1964)}]{hamilton_wd_jtb64a}
\bibinfo{author}{\bibfnamefont{W.~D.} \bibnamefont{Hamilton}},
  \bibinfo{journal}{J. Theor. Biol.} \textbf{\bibinfo{volume}{7}},
  \bibinfo{pages}{1} (\bibinfo{year}{1964}).

\bibitem[{\citenamefont{Alizon and Taylor}(2008)}]{Alizon2008}
\bibinfo{author}{\bibfnamefont{S.}~\bibnamefont{Alizon}} \bibnamefont{and}
  \bibinfo{author}{\bibfnamefont{P.}~\bibnamefont{Taylor}},
  \bibinfo{journal}{Evolution} \textbf{\bibinfo{volume}{62}},
  \bibinfo{pages}{1335} (\bibinfo{year}{2008}).

\bibitem[{\citenamefont{Sicardi et~al.}(2009)\citenamefont{Sicardi, Fort,
  Vainstein, and Arenzon}}]{Sicardi2009}
\bibinfo{author}{\bibfnamefont{E.~A.} \bibnamefont{Sicardi}},
  \bibinfo{author}{\bibfnamefont{H.}~\bibnamefont{Fort}},
  \bibinfo{author}{\bibfnamefont{M.~H.} \bibnamefont{Vainstein}},
  \bibnamefont{and} \bibinfo{author}{\bibfnamefont{J.~J.}
  \bibnamefont{Arenzon}}, \bibinfo{journal}{J. Theor. Biol.}
  \textbf{\bibinfo{volume}{256}}, \bibinfo{pages}{240} (\bibinfo{year}{2009}).

\bibitem[{\citenamefont{Trivers}(1971)}]{trivers_qrb71}
\bibinfo{author}{\bibfnamefont{R.~L.} \bibnamefont{Trivers}},
  \bibinfo{journal}{Q. Rev. Biol.} \textbf{\bibinfo{volume}{46}},
  \bibinfo{pages}{35} (\bibinfo{year}{1971}).

\bibitem[{\citenamefont{Axelrod and Hamilton}(1981)}]{axelrod_s81}
\bibinfo{author}{\bibfnamefont{R.}~\bibnamefont{Axelrod}} \bibnamefont{and}
  \bibinfo{author}{\bibfnamefont{W.~D.} \bibnamefont{Hamilton}},
  \bibinfo{journal}{Science} \textbf{\bibinfo{volume}{211}},
  \bibinfo{pages}{1390} (\bibinfo{year}{1981}).

\bibitem[{\citenamefont{Santos and Pacheco}(2005)}]{santos_prl05}
\bibinfo{author}{\bibfnamefont{F.~C.} \bibnamefont{Santos}} \bibnamefont{and}
  \bibinfo{author}{\bibfnamefont{J.~M.} \bibnamefont{Pacheco}},
  \bibinfo{journal}{Phys. Rev. Lett.} \textbf{\bibinfo{volume}{95}},
  \bibinfo{pages}{098104} (\bibinfo{year}{2005}).

\bibitem[{\citenamefont{Santos et~al.}(2006)\citenamefont{Santos, Pacheco, and
  Lenaerts}}]{santos_pnas06}
\bibinfo{author}{\bibfnamefont{F.~C.} \bibnamefont{Santos}},
  \bibinfo{author}{\bibfnamefont{J.~M.} \bibnamefont{Pacheco}},
  \bibnamefont{and} \bibinfo{author}{\bibfnamefont{T.}~\bibnamefont{Lenaerts}},
  \bibinfo{journal}{Proc. Natl. Acad. Sci. USA} \textbf{\bibinfo{volume}{103}},
  \bibinfo{pages}{3490} (\bibinfo{year}{2006}).

\bibitem[{\citenamefont{G{\'o}mez-Garde{\~n}es
  et~al.}(2007)\citenamefont{G{\'o}mez-Garde{\~n}es, Campillo, Flor{\'{\i}}a,
  and Moreno}}]{gomez-gardenes_prl07}
\bibinfo{author}{\bibfnamefont{J.}~\bibnamefont{G{\'o}mez-Garde{\~n}es}},
  \bibinfo{author}{\bibfnamefont{M.}~\bibnamefont{Campillo}},
  \bibinfo{author}{\bibfnamefont{L.~M.} \bibnamefont{Flor{\'{\i}}a}},
  \bibnamefont{and} \bibinfo{author}{\bibfnamefont{Y.}~\bibnamefont{Moreno}},
  \bibinfo{journal}{Phys. Rev. Lett.} \textbf{\bibinfo{volume}{98}},
  \bibinfo{pages}{108103} (\bibinfo{year}{2007}).

\bibitem[{\citenamefont{Wilson}(1977)}]{wilson_ds_an77}
\bibinfo{author}{\bibfnamefont{D.~S.} \bibnamefont{Wilson}},
  \bibinfo{journal}{Am. Nat.} \textbf{\bibinfo{volume}{111}},
  \bibinfo{pages}{157} (\bibinfo{year}{1977}).

\bibitem[{\citenamefont{Szolnoki and Szab{\'o}}(2007)}]{szolnoki_epl07}
\bibinfo{author}{\bibfnamefont{A.}~\bibnamefont{Szolnoki}} \bibnamefont{and}
  \bibinfo{author}{\bibfnamefont{G.}~\bibnamefont{Szab{\'o}}},
  \bibinfo{journal}{EPL} \textbf{\bibinfo{volume}{77}}, \bibinfo{pages}{30004}
  (\bibinfo{year}{2007}).

\bibitem[{\citenamefont{Perc and Szolnoki}(2008)}]{perc_pre08}
\bibinfo{author}{\bibfnamefont{M.}~\bibnamefont{Perc}} \bibnamefont{and}
  \bibinfo{author}{\bibfnamefont{A.}~\bibnamefont{Szolnoki}},
  \bibinfo{journal}{Phys. Rev. E} \textbf{\bibinfo{volume}{77}},
  \bibinfo{pages}{011904} (\bibinfo{year}{2008}).

\bibitem[{\citenamefont{Santos et~al.}(2008)\citenamefont{Santos, Santos, and
  Pacheco}}]{santos_n08}
\bibinfo{author}{\bibfnamefont{F.~C.} \bibnamefont{Santos}},
  \bibinfo{author}{\bibfnamefont{M.~D.} \bibnamefont{Santos}},
  \bibnamefont{and} \bibinfo{author}{\bibfnamefont{J.~M.}
  \bibnamefont{Pacheco}}, \bibinfo{journal}{Nature}
  \textbf{\bibinfo{volume}{454}}, \bibinfo{pages}{213} (\bibinfo{year}{2008}).

\bibitem[{\citenamefont{Santos et~al.}(2012)\citenamefont{Santos, Pinheiro,
  Lenaerts, and Pacheco}}]{santos_jtb12}
\bibinfo{author}{\bibfnamefont{F.~C.} \bibnamefont{Santos}},
  \bibinfo{author}{\bibfnamefont{F.}~\bibnamefont{Pinheiro}},
  \bibinfo{author}{\bibfnamefont{T.}~\bibnamefont{Lenaerts}}, \bibnamefont{and}
  \bibinfo{author}{\bibfnamefont{J.~M.} \bibnamefont{Pacheco}},
  \bibinfo{journal}{J. Theor. Biol.} \textbf{\bibinfo{volume}{299}},
  \bibinfo{pages}{88} (\bibinfo{year}{2012}).

\bibitem[{\citenamefont{Zimmermann et~al.}(2004)\citenamefont{Zimmermann,
  Egu{\'{\i}}luz, and San~Miguel}}]{zimmermann_pre04}
\bibinfo{author}{\bibfnamefont{M.~G.} \bibnamefont{Zimmermann}},
  \bibinfo{author}{\bibfnamefont{V.~M.} \bibnamefont{Egu{\'{\i}}luz}},
  \bibnamefont{and}
  \bibinfo{author}{\bibfnamefont{M.}~\bibnamefont{San~Miguel}},
  \bibinfo{journal}{Phys. Rev. E} \textbf{\bibinfo{volume}{69}},
  \bibinfo{pages}{065102(R)} (\bibinfo{year}{2004}).

\bibitem[{\citenamefont{Zimmermann and
  Egu{\'{\i}}luz}(2005)}]{zimmermann_pre05}
\bibinfo{author}{\bibfnamefont{M.~G.} \bibnamefont{Zimmermann}}
  \bibnamefont{and} \bibinfo{author}{\bibfnamefont{V.~M.}
  \bibnamefont{Egu{\'{\i}}luz}}, \bibinfo{journal}{Phys. Rev. E}
  \textbf{\bibinfo{volume}{72}}, \bibinfo{pages}{056118}
  (\bibinfo{year}{2005}).

\bibitem[{\citenamefont{Fu et~al.}(2009)\citenamefont{Fu, Wu, and
  Wang}}]{fu_pre09}
\bibinfo{author}{\bibfnamefont{F.}~\bibnamefont{Fu}},
  \bibinfo{author}{\bibfnamefont{T.}~\bibnamefont{Wu}}, \bibnamefont{and}
  \bibinfo{author}{\bibfnamefont{L.}~\bibnamefont{Wang}},
  \bibinfo{journal}{Phys. Rev. E} \textbf{\bibinfo{volume}{79}},
  \bibinfo{pages}{036101} (\bibinfo{year}{2009}).

\bibitem[{\citenamefont{Du et~al.}(2009)\citenamefont{Du, Cao, Hu, and
  Wang}}]{du_wb_epl09}
\bibinfo{author}{\bibfnamefont{W.-B.} \bibnamefont{Du}},
  \bibinfo{author}{\bibfnamefont{X.-B.} \bibnamefont{Cao}},
  \bibinfo{author}{\bibfnamefont{M.-B.} \bibnamefont{Hu}}, \bibnamefont{and}
  \bibinfo{author}{\bibfnamefont{W.-X.} \bibnamefont{Wang}},
  \bibinfo{journal}{EPL} \textbf{\bibinfo{volume}{87}}, \bibinfo{pages}{60004}
  (\bibinfo{year}{2009}).

\bibitem[{\citenamefont{Lee et~al.}(2011)\citenamefont{Lee, Holme, and
  Wu}}]{lee_s_prl11}
\bibinfo{author}{\bibfnamefont{S.}~\bibnamefont{Lee}},
  \bibinfo{author}{\bibfnamefont{P.}~\bibnamefont{Holme}}, \bibnamefont{and}
  \bibinfo{author}{\bibfnamefont{Z.-X.} \bibnamefont{Wu}},
  \bibinfo{journal}{Phys. Rev. Lett.} \textbf{\bibinfo{volume}{106}},
  \bibinfo{pages}{028702} (\bibinfo{year}{2011}).

\bibitem[{\citenamefont{G{\'o}mez-Garde{\~n}es
  et~al.}(2011)\citenamefont{G{\'o}mez-Garde{\~n}es, Vilone, and
  S{\'a}nchez}}]{gomez-gardenes_epl11}
\bibinfo{author}{\bibfnamefont{J.}~\bibnamefont{G{\'o}mez-Garde{\~n}es}},
  \bibinfo{author}{\bibfnamefont{D.}~\bibnamefont{Vilone}}, \bibnamefont{and}
  \bibinfo{author}{\bibfnamefont{A.}~\bibnamefont{S{\'a}nchez}},
  \bibinfo{journal}{EPL} \textbf{\bibinfo{volume}{95}}, \bibinfo{pages}{68003}
  (\bibinfo{year}{2011}).

\bibitem[{\citenamefont{Ohdaira and Terano}(2011)}]{ohdaira_jasss11}
\bibinfo{author}{\bibfnamefont{T.}~\bibnamefont{Ohdaira}} \bibnamefont{and}
  \bibinfo{author}{\bibfnamefont{T.}~\bibnamefont{Terano}},
  \bibinfo{journal}{Journal of Artificial Societies and Social Simulation}
  \textbf{\bibinfo{volume}{14}}, \bibinfo{pages}{3} (\bibinfo{year}{2011}).

\bibitem[{\citenamefont{Tanimoto et~al.}(2012)\citenamefont{Tanimoto, Brede,
  and Yamauchi}}]{tanimoto_pre12}
\bibinfo{author}{\bibfnamefont{J.}~\bibnamefont{Tanimoto}},
  \bibinfo{author}{\bibfnamefont{M.}~\bibnamefont{Brede}}, \bibnamefont{and}
  \bibinfo{author}{\bibfnamefont{A.}~\bibnamefont{Yamauchi}},
  \bibinfo{journal}{Phys. Rev. E} \textbf{\bibinfo{volume}{85}},
  \bibinfo{pages}{032101} (\bibinfo{year}{2012}).

\bibitem[{\citenamefont{Santos et~al.}(2014)\citenamefont{Santos, Dorogovtsev,
  and Mendes}}]{santos_md_srep14}
\bibinfo{author}{\bibfnamefont{M.}~\bibnamefont{Santos}},
  \bibinfo{author}{\bibfnamefont{S.~N.} \bibnamefont{Dorogovtsev}},
  \bibnamefont{and} \bibinfo{author}{\bibfnamefont{J.~F.~F.}
  \bibnamefont{Mendes}}, \bibinfo{journal}{Sci. Rep.}
  \textbf{\bibinfo{volume}{4}}, \bibinfo{pages}{4436} (\bibinfo{year}{2014}).

\bibitem[{\citenamefont{Pavlogiannis et~al.}(2015)\citenamefont{Pavlogiannis,
  Chatterjee, Adlam, and Nowak}}]{pavlogiannis_srep15}
\bibinfo{author}{\bibfnamefont{A.}~\bibnamefont{Pavlogiannis}},
  \bibinfo{author}{\bibfnamefont{K.}~\bibnamefont{Chatterjee}},
  \bibinfo{author}{\bibfnamefont{B.}~\bibnamefont{Adlam}}, \bibnamefont{and}
  \bibinfo{author}{\bibfnamefont{M.~A.} \bibnamefont{Nowak}},
  \bibinfo{journal}{Sci. Rep.} \textbf{\bibinfo{volume}{5}},
  \bibinfo{pages}{17147} (\bibinfo{year}{2015}).

\bibitem[{\citenamefont{Wu et~al.}(2015)\citenamefont{Wu, Rong, and
  Chen}}]{wu_zx_epl15}
\bibinfo{author}{\bibfnamefont{Z.-X.} \bibnamefont{Wu}},
  \bibinfo{author}{\bibfnamefont{Z.}~\bibnamefont{Rong}}, \bibnamefont{and}
  \bibinfo{author}{\bibfnamefont{M.~Z.~Q.} \bibnamefont{Chen}},
  \bibinfo{journal}{EPL} \textbf{\bibinfo{volume}{110}}, \bibinfo{pages}{30002}
  (\bibinfo{year}{2015}).

\bibitem[{\citenamefont{Hindersin and Traulsen}(2015)}]{hindersin_pcbi15}
\bibinfo{author}{\bibfnamefont{L.}~\bibnamefont{Hindersin}} \bibnamefont{and}
  \bibinfo{author}{\bibfnamefont{A.}~\bibnamefont{Traulsen}},
  \bibinfo{journal}{PLoS Comput. Biol.} \textbf{\bibinfo{volume}{11}},
  \bibinfo{pages}{e1004437} (\bibinfo{year}{2015}).

\bibitem[{\citenamefont{Chen et~al.}(2016)\citenamefont{Chen, Wu, Li, and
  Wang}}]{chen_w_pa16}
\bibinfo{author}{\bibfnamefont{W.}~\bibnamefont{Chen}},
  \bibinfo{author}{\bibfnamefont{T.}~\bibnamefont{Wu}},
  \bibinfo{author}{\bibfnamefont{Z.}~\bibnamefont{Li}}, \bibnamefont{and}
  \bibinfo{author}{\bibfnamefont{L.}~\bibnamefont{Wang}},
  \bibinfo{journal}{Physica A} \textbf{\bibinfo{volume}{443}},
  \bibinfo{pages}{192} (\bibinfo{year}{2016}).

\bibitem[{\citenamefont{Szab{\'o} and F{\'a}th}(2007)}]{szabo_pr07}
\bibinfo{author}{\bibfnamefont{G.}~\bibnamefont{Szab{\'o}}} \bibnamefont{and}
  \bibinfo{author}{\bibfnamefont{G.}~\bibnamefont{F{\'a}th}},
  \bibinfo{journal}{Phys. Rep.} \textbf{\bibinfo{volume}{446}},
  \bibinfo{pages}{97} (\bibinfo{year}{2007}).

\bibitem[{\citenamefont{Roca et~al.}(2009{\natexlab{a}})\citenamefont{Roca,
  Cuesta, and S{\'a}nchez}}]{roca_plr09}
\bibinfo{author}{\bibfnamefont{C.~P.} \bibnamefont{Roca}},
  \bibinfo{author}{\bibfnamefont{J.~A.} \bibnamefont{Cuesta}},
  \bibnamefont{and}
  \bibinfo{author}{\bibfnamefont{A.}~\bibnamefont{S{\'a}nchez}},
  \bibinfo{journal}{Phys. Life Rev.} \textbf{\bibinfo{volume}{6}},
  \bibinfo{pages}{208} (\bibinfo{year}{2009}{\natexlab{a}}).

\bibitem[{\citenamefont{Perc and Szolnoki}(2010)}]{perc_bs10}
\bibinfo{author}{\bibfnamefont{M.}~\bibnamefont{Perc}} \bibnamefont{and}
  \bibinfo{author}{\bibfnamefont{A.}~\bibnamefont{Szolnoki}},
  \bibinfo{journal}{BioSystems} \textbf{\bibinfo{volume}{99}},
  \bibinfo{pages}{109} (\bibinfo{year}{2010}).

\bibitem[{\citenamefont{Perc et~al.}(2013)\citenamefont{Perc,
  G{\'o}mez-Garde{\~n}es, Szolnoki, and Flor{\'{\i}a and Y.
  Moreno}}}]{perc_jrsi13}
\bibinfo{author}{\bibfnamefont{M.}~\bibnamefont{Perc}},
  \bibinfo{author}{\bibfnamefont{J.}~\bibnamefont{G{\'o}mez-Garde{\~n}es}},
  \bibinfo{author}{\bibfnamefont{A.}~\bibnamefont{Szolnoki}}, \bibnamefont{and}
  \bibinfo{author}{\bibfnamefont{L.~M.} \bibnamefont{Flor{\'{\i}a and Y.
  Moreno}}}, \bibinfo{journal}{J. R. Soc. Interface}
  \textbf{\bibinfo{volume}{10}}, \bibinfo{pages}{20120997}
  (\bibinfo{year}{2013}).

\bibitem[{\citenamefont{Pacheco et~al.}(2014)\citenamefont{Pacheco,
  Vasconcelos, and Santos}}]{pacheco_plrev14}
\bibinfo{author}{\bibfnamefont{J.~M.} \bibnamefont{Pacheco}},
  \bibinfo{author}{\bibfnamefont{V.~V.} \bibnamefont{Vasconcelos}},
  \bibnamefont{and} \bibinfo{author}{\bibfnamefont{F.~C.}
  \bibnamefont{Santos}}, \bibinfo{journal}{Physics of Life Reviews}
  \textbf{\bibinfo{volume}{11}}, \bibinfo{pages}{573} (\bibinfo{year}{2014}).

\bibitem[{\citenamefont{Wang et~al.}(2015{\natexlab{a}})\citenamefont{Wang,
  Wang, Szolnoki, and Perc}}]{wang_z_epjb15}
\bibinfo{author}{\bibfnamefont{Z.}~\bibnamefont{Wang}},
  \bibinfo{author}{\bibfnamefont{L.}~\bibnamefont{Wang}},
  \bibinfo{author}{\bibfnamefont{A.}~\bibnamefont{Szolnoki}}, \bibnamefont{and}
  \bibinfo{author}{\bibfnamefont{M.}~\bibnamefont{Perc}},
  \bibinfo{journal}{Eur. Phys. J. B} \textbf{\bibinfo{volume}{88}},
  \bibinfo{pages}{124} (\bibinfo{year}{2015}{\natexlab{a}}).

\bibitem[{\citenamefont{Wang et~al.}(2015{\natexlab{b}})\citenamefont{Wang,
  Kokubo, Jusup, and Tanimoto}}]{wang2015universal}
\bibinfo{author}{\bibfnamefont{Z.}~\bibnamefont{Wang}},
  \bibinfo{author}{\bibfnamefont{S.}~\bibnamefont{Kokubo}},
  \bibinfo{author}{\bibfnamefont{M.}~\bibnamefont{Jusup}}, \bibnamefont{and}
  \bibinfo{author}{\bibfnamefont{J.}~\bibnamefont{Tanimoto}},
  \bibinfo{journal}{Phys. Life Rev.} \textbf{\bibinfo{volume}{14}},
  \bibinfo{pages}{1} (\bibinfo{year}{2015}{\natexlab{b}}).

\bibitem[{\citenamefont{Vukov et~al.}(2006)\citenamefont{Vukov, Szab{\'o}, and
  Szolnoki}}]{vukov_pre06}
\bibinfo{author}{\bibfnamefont{J.}~\bibnamefont{Vukov}},
  \bibinfo{author}{\bibfnamefont{G.}~\bibnamefont{Szab{\'o}}},
  \bibnamefont{and} \bibinfo{author}{\bibfnamefont{A.}~\bibnamefont{Szolnoki}},
  \bibinfo{journal}{Phys. Rev. E} \textbf{\bibinfo{volume}{73}},
  \bibinfo{pages}{067103} (\bibinfo{year}{2006}).

\bibitem[{\citenamefont{Perc}(2006)}]{perc_njp06a}
\bibinfo{author}{\bibfnamefont{M.}~\bibnamefont{Perc}}, \bibinfo{journal}{New
  J. Phys.} \textbf{\bibinfo{volume}{8}}, \bibinfo{pages}{22}
  (\bibinfo{year}{2006}).

\bibitem[{\citenamefont{Tanimoto}(2007)}]{tanimoto_pre07b}
\bibinfo{author}{\bibfnamefont{J.}~\bibnamefont{Tanimoto}},
  \bibinfo{journal}{Phys. Rev. E} \textbf{\bibinfo{volume}{76}},
  \bibinfo{pages}{041130} (\bibinfo{year}{2007}).

\bibitem[{\citenamefont{Szolnoki
  et~al.}(2008{\natexlab{a}})\citenamefont{Szolnoki, Perc, and
  Danku}}]{szolnoki_epl08}
\bibinfo{author}{\bibfnamefont{A.}~\bibnamefont{Szolnoki}},
  \bibinfo{author}{\bibfnamefont{M.}~\bibnamefont{Perc}}, \bibnamefont{and}
  \bibinfo{author}{\bibfnamefont{Z.}~\bibnamefont{Danku}},
  \bibinfo{journal}{EPL} \textbf{\bibinfo{volume}{84}}, \bibinfo{pages}{50007}
  (\bibinfo{year}{2008}{\natexlab{a}}).

\bibitem[{\citenamefont{Szolnoki
  et~al.}(2008{\natexlab{b}})\citenamefont{Szolnoki, Perc, and
  Szab{\'o}}}]{szolnoki_epjb08}
\bibinfo{author}{\bibfnamefont{A.}~\bibnamefont{Szolnoki}},
  \bibinfo{author}{\bibfnamefont{M.}~\bibnamefont{Perc}}, \bibnamefont{and}
  \bibinfo{author}{\bibfnamefont{G.}~\bibnamefont{Szab{\'o}}},
  \bibinfo{journal}{Eur. Phys. J. B} \textbf{\bibinfo{volume}{61}},
  \bibinfo{pages}{505} (\bibinfo{year}{2008}{\natexlab{b}}).

\bibitem[{\citenamefont{Jiang et~al.}(2009)\citenamefont{Jiang, Zhao, Yang,
  Wakeling, Wang, and Zhou}}]{jiang_ll_pre09}
\bibinfo{author}{\bibfnamefont{L.-L.} \bibnamefont{Jiang}},
  \bibinfo{author}{\bibfnamefont{M.}~\bibnamefont{Zhao}},
  \bibinfo{author}{\bibfnamefont{H.-X.} \bibnamefont{Yang}},
  \bibinfo{author}{\bibfnamefont{J.}~\bibnamefont{Wakeling}},
  \bibinfo{author}{\bibfnamefont{B.-H.} \bibnamefont{Wang}}, \bibnamefont{and}
  \bibinfo{author}{\bibfnamefont{T.}~\bibnamefont{Zhou}},
  \bibinfo{journal}{Phys. Rev. E} \textbf{\bibinfo{volume}{80}},
  \bibinfo{pages}{031144} (\bibinfo{year}{2009}).

\bibitem[{\citenamefont{Devlin and Treloar}(2009)}]{devlin_pre09}
\bibinfo{author}{\bibfnamefont{S.}~\bibnamefont{Devlin}} \bibnamefont{and}
  \bibinfo{author}{\bibfnamefont{T.}~\bibnamefont{Treloar}},
  \bibinfo{journal}{Phys. Rev. E} \textbf{\bibinfo{volume}{79}},
  \bibinfo{pages}{016107} (\bibinfo{year}{2009}).

\bibitem[{\citenamefont{Shigaki et~al.}(2012)\citenamefont{Shigaki, Kokubo,
  Tanimoto, Hagishima, and Ikegaya}}]{shigaki_epl12}
\bibinfo{author}{\bibfnamefont{K.}~\bibnamefont{Shigaki}},
  \bibinfo{author}{\bibfnamefont{S.}~\bibnamefont{Kokubo}},
  \bibinfo{author}{\bibfnamefont{J.}~\bibnamefont{Tanimoto}},
  \bibinfo{author}{\bibfnamefont{A.}~\bibnamefont{Hagishima}},
  \bibnamefont{and} \bibinfo{author}{\bibfnamefont{N.}~\bibnamefont{Ikegaya}},
  \bibinfo{journal}{EPL} \textbf{\bibinfo{volume}{98}}, \bibinfo{pages}{40008}
  (\bibinfo{year}{2012}).

\bibitem[{\citenamefont{Hauser et~al.}(2014)\citenamefont{Hauser, Traulsen, and
  Nowak}}]{hauser_jtb14}
\bibinfo{author}{\bibfnamefont{O.~P.} \bibnamefont{Hauser}},
  \bibinfo{author}{\bibfnamefont{A.}~\bibnamefont{Traulsen}}, \bibnamefont{and}
  \bibinfo{author}{\bibfnamefont{M.~A.} \bibnamefont{Nowak}},
  \bibinfo{journal}{J. Theor. Biol.} \textbf{\bibinfo{volume}{343}},
  \bibinfo{pages}{178} (\bibinfo{year}{2014}).

\bibitem[{\citenamefont{Yuan and Xia}(20141)}]{yuan_wj_pone14}
\bibinfo{author}{\bibfnamefont{W.-J.} \bibnamefont{Yuan}} \bibnamefont{and}
  \bibinfo{author}{\bibfnamefont{C.-Y.} \bibnamefont{Xia}},
  \bibinfo{journal}{PLoS ONE} \textbf{\bibinfo{volume}{9}},
  \bibinfo{pages}{e91012} (\bibinfo{year}{20141}).

\bibitem[{\citenamefont{Iwata and Akiyama}(2015)}]{iwa_pha15}
\bibinfo{author}{\bibfnamefont{M.}~\bibnamefont{Iwata}} \bibnamefont{and}
  \bibinfo{author}{\bibfnamefont{E.}~\bibnamefont{Akiyama}},
  \bibinfo{journal}{Physica A} \textbf{\bibinfo{volume}{448}},
  \bibinfo{pages}{224} (\bibinfo{year}{2015}).

\bibitem[{\citenamefont{Amaral et~al.}(2015)\citenamefont{Amaral, Wardil, and
  da~Silva}}]{amaral_jpa15}
\bibinfo{author}{\bibfnamefont{M.~A.} \bibnamefont{Amaral}},
  \bibinfo{author}{\bibfnamefont{L.}~\bibnamefont{Wardil}}, \bibnamefont{and}
  \bibinfo{author}{\bibfnamefont{J.~K.~L.} \bibnamefont{da~Silva}},
  \bibinfo{journal}{J. Phys. A} \textbf{\bibinfo{volume}{48}},
  \bibinfo{pages}{445002} (\bibinfo{year}{2015}).

\bibitem[{\citenamefont{Tanimoto and Kishimoto}(2015)}]{tanimoto2015network}
\bibinfo{author}{\bibfnamefont{J.}~\bibnamefont{Tanimoto}} \bibnamefont{and}
  \bibinfo{author}{\bibfnamefont{N.}~\bibnamefont{Kishimoto}},
  \bibinfo{journal}{Phys. Rev. E} \textbf{\bibinfo{volume}{91}},
  \bibinfo{pages}{042106} (\bibinfo{year}{2015}).

\bibitem[{\citenamefont{Liu et~al.}(2015)\citenamefont{Liu, Jia, and
  Rong}}]{liu_rr_epl15}
\bibinfo{author}{\bibfnamefont{R.-R.} \bibnamefont{Liu}},
  \bibinfo{author}{\bibfnamefont{C.-X.} \bibnamefont{Jia}}, \bibnamefont{and}
  \bibinfo{author}{\bibfnamefont{Z.}~\bibnamefont{Rong}},
  \bibinfo{journal}{EPL} \textbf{\bibinfo{volume}{112}}, \bibinfo{pages}{48005}
  (\bibinfo{year}{2015}).

\bibitem[{\citenamefont{Javarone}(2016)}]{javarone_epjb16}
\bibinfo{author}{\bibfnamefont{M.~A.} \bibnamefont{Javarone}},
  \bibinfo{journal}{Eur. Phys. J. B} \textbf{\bibinfo{volume}{89}},
  \bibinfo{pages}{42} (\bibinfo{year}{2016}).

\bibitem[{\citenamefont{Amaral et~al.}(2016)\citenamefont{Amaral, Wardil, Perc,
  and da~Silva}}]{amaral_pre16}
\bibinfo{author}{\bibfnamefont{M.~A.} \bibnamefont{Amaral}},
  \bibinfo{author}{\bibfnamefont{L.}~\bibnamefont{Wardil}},
  \bibinfo{author}{\bibfnamefont{M.}~\bibnamefont{Perc}}, \bibnamefont{and}
  \bibinfo{author}{\bibfnamefont{J.~K.L.} \bibnamefont{da~Silva}},
  \bibinfo{journal}{Phys. Rev. E} \textbf{\bibinfo{volume}{93}},
  \bibinfo{pages}{042304} (\bibinfo{year}{2016}).

\bibitem[{\citenamefont{Matsuzawa et~al.}(2016)\citenamefont{Matsuzawa,
  Tanimoto, and Fukuda}}]{matsuzawa2016spatial}
\bibinfo{author}{\bibfnamefont{R.}~\bibnamefont{Matsuzawa}},
  \bibinfo{author}{\bibfnamefont{J.}~\bibnamefont{Tanimoto}}, \bibnamefont{and}
  \bibinfo{author}{\bibfnamefont{E.}~\bibnamefont{Fukuda}},
  \bibinfo{journal}{Phys. Rev. E} \textbf{\bibinfo{volume}{94}},
  \bibinfo{pages}{022114} (\bibinfo{year}{2016}).

\bibitem[{\citenamefont{Javarone and Battiston}(2016)}]{javarone2016role}
\bibinfo{author}{\bibfnamefont{M.~A.} \bibnamefont{Javarone}} \bibnamefont{and}
  \bibinfo{author}{\bibfnamefont{F.}~\bibnamefont{Battiston}},
  \bibinfo{journal}{J. Stat. Mech.} \textbf{\bibinfo{volume}{2016}},
  \bibinfo{pages}{073404} (\bibinfo{year}{2016}).

\bibitem[{\citenamefont{Javarone et~al.}(2016)\citenamefont{Javarone,
  Antonioni, and Caravelli}}]{javarone2016conformity}
\bibinfo{author}{\bibfnamefont{M.~A.} \bibnamefont{Javarone}},
  \bibinfo{author}{\bibfnamefont{A.}~\bibnamefont{Antonioni}},
  \bibnamefont{and}
  \bibinfo{author}{\bibfnamefont{F.}~\bibnamefont{Caravelli}},
  \bibinfo{journal}{EPL} \textbf{\bibinfo{volume}{114}}, \bibinfo{pages}{38001}
  (\bibinfo{year}{2016}).

\bibitem[{\citenamefont{Wardil and da~Silva}(2009)}]{wardil_epl09}
\bibinfo{author}{\bibfnamefont{L.}~\bibnamefont{Wardil}} \bibnamefont{and}
  \bibinfo{author}{\bibfnamefont{J.~K.~L.} \bibnamefont{da~Silva}},
  \bibinfo{journal}{EPL} \textbf{\bibinfo{volume}{86}}, \bibinfo{pages}{38001}
  (\bibinfo{year}{2009}).

\bibitem[{\citenamefont{Robinson et~al.}(2011)\citenamefont{Robinson, Feldman,
  and McKay}}]{Robinson2011}
\bibinfo{author}{\bibfnamefont{M.~D.} \bibnamefont{Robinson}},
  \bibinfo{author}{\bibfnamefont{D.~P.} \bibnamefont{Feldman}},
  \bibnamefont{and} \bibinfo{author}{\bibfnamefont{S.~R.} \bibnamefont{McKay}},
  \bibinfo{journal}{Chaos} \textbf{\bibinfo{volume}{21}},
  \bibinfo{pages}{037114} (\bibinfo{year}{2011}).

\bibitem[{\citenamefont{Binder and Landau}(1980)}]{binder_prb80}
\bibinfo{author}{\bibfnamefont{K.}~\bibnamefont{Binder}} \bibnamefont{and}
  \bibinfo{author}{\bibfnamefont{D.~P.} \bibnamefont{Landau}},
  \bibinfo{journal}{Phys. Rev. B} \textbf{\bibinfo{volume}{21}},
  \bibinfo{pages}{1941} (\bibinfo{year}{1980}).

\bibitem[{\citenamefont{Gracia-L{\'a}zaro
  et~al.}(2012{\natexlab{a}})\citenamefont{Gracia-L{\'a}zaro, Cuesta,
  S{\'a}nchez, and Moreno}}]{gracia-lazaro_srep12}
\bibinfo{author}{\bibfnamefont{C.}~\bibnamefont{Gracia-L{\'a}zaro}},
  \bibinfo{author}{\bibfnamefont{J.}~\bibnamefont{Cuesta}},
  \bibinfo{author}{\bibfnamefont{A.}~\bibnamefont{S{\'a}nchez}},
  \bibnamefont{and} \bibinfo{author}{\bibfnamefont{Y.}~\bibnamefont{Moreno}},
  \bibinfo{journal}{Sci. Rep.} \textbf{\bibinfo{volume}{2}},
  \bibinfo{pages}{325} (\bibinfo{year}{2012}{\natexlab{a}}).

\bibitem[{\citenamefont{Gracia-L{\'a}zaro
  et~al.}(2012{\natexlab{b}})\citenamefont{Gracia-L{\'a}zaro, Ferrer, Ruiz,
  Taranc{\'o}n, Cuesta, S{\'a}nchez, and Moreno}}]{gracia-lazaro_pnas12}
\bibinfo{author}{\bibfnamefont{C.}~\bibnamefont{Gracia-L{\'a}zaro}},
  \bibinfo{author}{\bibfnamefont{A.}~\bibnamefont{Ferrer}},
  \bibinfo{author}{\bibfnamefont{G.}~\bibnamefont{Ruiz}},
  \bibinfo{author}{\bibfnamefont{A.}~\bibnamefont{Taranc{\'o}n}},
  \bibinfo{author}{\bibfnamefont{J.}~\bibnamefont{Cuesta}},
  \bibinfo{author}{\bibfnamefont{A.}~\bibnamefont{S{\'a}nchez}},
  \bibnamefont{and} \bibinfo{author}{\bibfnamefont{Y.}~\bibnamefont{Moreno}},
  \bibinfo{journal}{Proc. Natl. Acad. Sci. USA} \textbf{\bibinfo{volume}{109}},
  \bibinfo{pages}{12922} (\bibinfo{year}{2012}{\natexlab{b}}).

\bibitem[{\citenamefont{Gruji{\'{c}} et~al.}(2014)\citenamefont{Gruji{\'{c}},
  Gracia-L{\'{a}}zaro, Milinski, Semmann, Traulsen, Cuesta, Moreno, and
  S{\'{a}}nchez}}]{Grujic2014}
\bibinfo{author}{\bibfnamefont{J.}~\bibnamefont{Gruji{\'{c}}}},
  \bibinfo{author}{\bibfnamefont{C.}~\bibnamefont{Gracia-L{\'{a}}zaro}},
  \bibinfo{author}{\bibfnamefont{M.}~\bibnamefont{Milinski}},
  \bibinfo{author}{\bibfnamefont{D.}~\bibnamefont{Semmann}},
  \bibinfo{author}{\bibfnamefont{A.}~\bibnamefont{Traulsen}},
  \bibinfo{author}{\bibfnamefont{J.~A.} \bibnamefont{Cuesta}},
  \bibinfo{author}{\bibfnamefont{Y.}~\bibnamefont{Moreno}}, \bibnamefont{and}
  \bibinfo{author}{\bibfnamefont{A.}~\bibnamefont{S{\'{a}}nchez}},
  \bibinfo{journal}{Sci. Rep.} \textbf{\bibinfo{volume}{4}},
  \bibinfo{pages}{4615} (\bibinfo{year}{2014}).

\bibitem[{\citenamefont{Vukov et~al.}(2012)\citenamefont{Vukov, Santos, and
  Pacheco}}]{vukov_njp12}
\bibinfo{author}{\bibfnamefont{J.}~\bibnamefont{Vukov}},
  \bibinfo{author}{\bibfnamefont{F.}~\bibnamefont{Santos}}, \bibnamefont{and}
  \bibinfo{author}{\bibfnamefont{J.}~\bibnamefont{Pacheco}},
  \bibinfo{journal}{New J. Phys.} \textbf{\bibinfo{volume}{14}},
  \bibinfo{pages}{063031} (\bibinfo{year}{2012}).

\bibitem[{\citenamefont{Blume and Gneezy}(2010)}]{blume_a_geb10}
\bibinfo{author}{\bibfnamefont{A.}~\bibnamefont{Blume}} \bibnamefont{and}
  \bibinfo{author}{\bibfnamefont{U.}~\bibnamefont{Gneezy}},
  \bibinfo{journal}{Games Econ. Behav.}  (\bibinfo{year}{2010}).

\bibitem[{\citenamefont{Bonawitz et~al.}(2014)\citenamefont{Bonawitz, Denison,
  Gopnik, and Griffiths}}]{bonawitz_cg14}
\bibinfo{author}{\bibfnamefont{E.}~\bibnamefont{Bonawitz}},
  \bibinfo{author}{\bibfnamefont{S.}~\bibnamefont{Denison}},
  \bibinfo{author}{\bibfnamefont{A.}~\bibnamefont{Gopnik}}, \bibnamefont{and}
  \bibinfo{author}{\bibfnamefont{T.~L.} \bibnamefont{Griffiths}},
  \bibinfo{journal}{Cogn. Psychol.} \textbf{\bibinfo{volume}{74}},
  \bibinfo{pages}{35 } (\bibinfo{year}{2014}).

\bibitem[{\citenamefont{Blume}(1995)}]{blume_l_geb95}
\bibinfo{author}{\bibfnamefont{L.~E.} \bibnamefont{Blume}},
  \bibinfo{journal}{Games Econ. Behav.} \textbf{\bibinfo{volume}{11}},
  \bibinfo{pages}{111} (\bibinfo{year}{1995}).

\bibitem[{\citenamefont{Szab{\'o} et~al.}(2013)\citenamefont{Szab{\'o},
  Szolnoki, and Czak{\'o}}}]{szabo_jtb12b}
\bibinfo{author}{\bibfnamefont{G.}~\bibnamefont{Szab{\'o}}},
  \bibinfo{author}{\bibfnamefont{A.}~\bibnamefont{Szolnoki}}, \bibnamefont{and}
  \bibinfo{author}{\bibfnamefont{L.}~\bibnamefont{Czak{\'o}}},
  \bibinfo{journal}{J. Theor. Biol.} \textbf{\bibinfo{volume}{317}},
  \bibinfo{pages}{126} (\bibinfo{year}{2013}).

\bibitem[{\citenamefont{Szab{\'o} et~al.}(2005)\citenamefont{Szab{\'o}, Vukov,
  and Szolnoki}}]{szabo_pre05}
\bibinfo{author}{\bibfnamefont{G.}~\bibnamefont{Szab{\'o}}},
  \bibinfo{author}{\bibfnamefont{J.}~\bibnamefont{Vukov}}, \bibnamefont{and}
  \bibinfo{author}{\bibfnamefont{A.}~\bibnamefont{Szolnoki}},
  \bibinfo{journal}{Phys. Rev. E} \textbf{\bibinfo{volume}{72}},
  \bibinfo{pages}{047107} (\bibinfo{year}{2005}).

\bibitem[{\citenamefont{Wedekind and Milinski}(1996)}]{wedekind_pnas96}
\bibinfo{author}{\bibfnamefont{C.}~\bibnamefont{Wedekind}} \bibnamefont{and}
  \bibinfo{author}{\bibfnamefont{M.}~\bibnamefont{Milinski}},
  \bibinfo{journal}{Proc. Natl. Acad. Sci. USA} \textbf{\bibinfo{volume}{93}},
  \bibinfo{pages}{2686} (\bibinfo{year}{1996}).

\bibitem[{\citenamefont{Dalton}(2010)}]{Dalton2010}
\bibinfo{author}{\bibfnamefont{P.~S.} \bibnamefont{Dalton}},
  \bibinfo{journal}{SSRN Electron. J.} \textbf{\bibinfo{volume}{2010-23}},
  \bibinfo{pages}{1} (\bibinfo{year}{2010}).

\bibitem[{\citenamefont{Macy and Flache}(2002)}]{macy_pnas02}
\bibinfo{author}{\bibfnamefont{M.~W.} \bibnamefont{Macy}} \bibnamefont{and}
  \bibinfo{author}{\bibfnamefont{A.}~\bibnamefont{Flache}},
  \bibinfo{journal}{Proc. Natl. Acad. Sci. USA} \textbf{\bibinfo{volume}{99}},
  \bibinfo{pages}{7229} (\bibinfo{year}{2002}).

\bibitem[{\citenamefont{Roca et~al.}(2009{\natexlab{b}})\citenamefont{Roca,
  Cuesta, and S{\'a}nchez}}]{roca_epjb09}
\bibinfo{author}{\bibfnamefont{C.~P.} \bibnamefont{Roca}},
  \bibinfo{author}{\bibfnamefont{J.~A.} \bibnamefont{Cuesta}},
  \bibnamefont{and}
  \bibinfo{author}{\bibfnamefont{A.}~\bibnamefont{S{\'a}nchez}},
  \bibinfo{journal}{Eur. Phys. J. B} \textbf{\bibinfo{volume}{71}},
  \bibinfo{pages}{587} (\bibinfo{year}{2009}{\natexlab{b}}).

\bibitem[{\citenamefont{Sysi-Aho et~al.}(2005)\citenamefont{Sysi-Aho,
  Saram{\"a}ki, Kert{\'e}sz, and Kaski}}]{sysiaho_epjb05}
\bibinfo{author}{\bibfnamefont{M.}~\bibnamefont{Sysi-Aho}},
  \bibinfo{author}{\bibfnamefont{J.}~\bibnamefont{Saram{\"a}ki}},
  \bibinfo{author}{\bibfnamefont{J.}~\bibnamefont{Kert{\'e}sz}},
  \bibnamefont{and} \bibinfo{author}{\bibfnamefont{K.}~\bibnamefont{Kaski}},
  \bibinfo{journal}{Eur. Phys. J. B} \textbf{\bibinfo{volume}{44}},
  \bibinfo{pages}{129} (\bibinfo{year}{2005}).

\bibitem[{\citenamefont{Gruji{\'c} et~al.}(2010)\citenamefont{Gruji{\'c},
  Fosco, Araujo, Cuesta, and S{\'a}nchez}}]{grujic_pone10}
\bibinfo{author}{\bibfnamefont{J.}~\bibnamefont{Gruji{\'c}}},
  \bibinfo{author}{\bibfnamefont{C.}~\bibnamefont{Fosco}},
  \bibinfo{author}{\bibfnamefont{L.}~\bibnamefont{Araujo}},
  \bibinfo{author}{\bibfnamefont{J.~A.} \bibnamefont{Cuesta}},
  \bibnamefont{and}
  \bibinfo{author}{\bibfnamefont{A.}~\bibnamefont{S{\'a}nchez}},
  \bibinfo{journal}{PLoS ONE} \textbf{\bibinfo{volume}{5}},
  \bibinfo{pages}{e13749} (\bibinfo{year}{2010}).

\bibitem[{\citenamefont{Szolnoki et~al.}(2011)\citenamefont{Szolnoki, Xie,
  Wang, and Perc}}]{szolnoki_epl11}
\bibinfo{author}{\bibfnamefont{A.}~\bibnamefont{Szolnoki}},
  \bibinfo{author}{\bibfnamefont{N.-G.} \bibnamefont{Xie}},
  \bibinfo{author}{\bibfnamefont{C.}~\bibnamefont{Wang}}, \bibnamefont{and}
  \bibinfo{author}{\bibfnamefont{M.}~\bibnamefont{Perc}},
  \bibinfo{journal}{EPL} \textbf{\bibinfo{volume}{96}}, \bibinfo{pages}{38002}
  (\bibinfo{year}{2011}).

\bibitem[{\citenamefont{Roca et~al.}(2009{\natexlab{c}})\citenamefont{Roca,
  Cuesta, and S{\'a}nchez}}]{roca_pre09}
\bibinfo{author}{\bibfnamefont{C.~P.} \bibnamefont{Roca}},
  \bibinfo{author}{\bibfnamefont{J.~A.} \bibnamefont{Cuesta}},
  \bibnamefont{and}
  \bibinfo{author}{\bibfnamefont{A.}~\bibnamefont{S{\'a}nchez}},
  \bibinfo{journal}{Phys. Rev. E} \textbf{\bibinfo{volume}{80}},
  \bibinfo{pages}{046106} (\bibinfo{year}{2009}{\natexlab{c}}).

\bibitem[{\citenamefont{Szab{\'o} et~al.}(2010)\citenamefont{Szab{\'o},
  Szolnoki, Varga, and Hanusovszky}}]{szabo_pre10}
\bibinfo{author}{\bibfnamefont{G.}~\bibnamefont{Szab{\'o}}},
  \bibinfo{author}{\bibfnamefont{A.}~\bibnamefont{Szolnoki}},
  \bibinfo{author}{\bibfnamefont{M.}~\bibnamefont{Varga}}, \bibnamefont{and}
  \bibinfo{author}{\bibfnamefont{L.}~\bibnamefont{Hanusovszky}},
  \bibinfo{journal}{Phys. Rev. E} \textbf{\bibinfo{volume}{82}},
  \bibinfo{pages}{026110} (\bibinfo{year}{2010}).

\bibitem[{\citenamefont{Szab{\'o} and Szolnoki}(2012)}]{szabo_jtb12}
\bibinfo{author}{\bibfnamefont{G.}~\bibnamefont{Szab{\'o}}} \bibnamefont{and}
  \bibinfo{author}{\bibfnamefont{A.}~\bibnamefont{Szolnoki}},
  \bibinfo{journal}{J. Theor. Biol.} \textbf{\bibinfo{volume}{299}},
  \bibinfo{pages}{81} (\bibinfo{year}{2012}).

\bibitem[{\citenamefont{Wang et~al.}(2012)\citenamefont{Wang, Szolnoki, and
  Perc}}]{wang_z_srep12}
\bibinfo{author}{\bibfnamefont{Z.}~\bibnamefont{Wang}},
  \bibinfo{author}{\bibfnamefont{A.}~\bibnamefont{Szolnoki}}, \bibnamefont{and}
  \bibinfo{author}{\bibfnamefont{M.}~\bibnamefont{Perc}},
  \bibinfo{journal}{Sci. Rep.} \textbf{\bibinfo{volume}{2}},
  \bibinfo{pages}{369} (\bibinfo{year}{2012}).

\bibitem[{\citenamefont{Glauber}(1963)}]{glauber_jmp63}
\bibinfo{author}{\bibfnamefont{R.~J.} \bibnamefont{Glauber}},
  \bibinfo{journal}{J. Math. Phys} \textbf{\bibinfo{volume}{4}},
  \bibinfo{pages}{294} (\bibinfo{year}{1963}).

\bibitem[{\citenamefont{Hauert and Szab{\'o}}(2005)}]{hauert_ajp05}
\bibinfo{author}{\bibfnamefont{C.}~\bibnamefont{Hauert}} \bibnamefont{and}
  \bibinfo{author}{\bibfnamefont{G.}~\bibnamefont{Szab{\'o}}},
  \bibinfo{journal}{Am. J. Phys.} \textbf{\bibinfo{volume}{73}},
  \bibinfo{pages}{405} (\bibinfo{year}{2005}).

\bibitem[{\citenamefont{Hauert and Doebeli}(2004)}]{hauert_n04}
\bibinfo{author}{\bibfnamefont{C.}~\bibnamefont{Hauert}} \bibnamefont{and}
  \bibinfo{author}{\bibfnamefont{M.}~\bibnamefont{Doebeli}},
  \bibinfo{journal}{Nature} \textbf{\bibinfo{volume}{428}},
  \bibinfo{pages}{643} (\bibinfo{year}{2004}).

\bibitem[{\citenamefont{Szab{\'o} et~al.}(2007)\citenamefont{Szab{\'o},
  Szolnoki, and Sznaider}}]{szabo_pre07}
\bibinfo{author}{\bibfnamefont{G.}~\bibnamefont{Szab{\'o}}},
  \bibinfo{author}{\bibfnamefont{A.}~\bibnamefont{Szolnoki}}, \bibnamefont{and}
  \bibinfo{author}{\bibfnamefont{G.~A.} \bibnamefont{Sznaider}},
  \bibinfo{journal}{Phys. Rev. E} \textbf{\bibinfo{volume}{76}},
  \bibinfo{pages}{051921} (\bibinfo{year}{2007}).

\bibitem[{\citenamefont{Nowak and Sigmund}(2004)}]{nowak_s04}
\bibinfo{author}{\bibfnamefont{M.~A.} \bibnamefont{Nowak}} \bibnamefont{and}
  \bibinfo{author}{\bibfnamefont{K.}~\bibnamefont{Sigmund}},
  \bibinfo{journal}{Science} \textbf{\bibinfo{volume}{303}},
  \bibinfo{pages}{793} (\bibinfo{year}{2004}).

\bibitem[{\citenamefont{Choi et~al.}(2015)\citenamefont{Choi, Yook, and
  Kim}}]{Choi2015}
\bibinfo{author}{\bibfnamefont{W.}~\bibnamefont{Choi}},
  \bibinfo{author}{\bibfnamefont{S.-H.} \bibnamefont{Yook}}, \bibnamefont{and}
  \bibinfo{author}{\bibfnamefont{Y.}~\bibnamefont{Kim}},
  \bibinfo{journal}{Phys. Rev. E} \textbf{\bibinfo{volume}{92}},
  \bibinfo{pages}{052140} (\bibinfo{year}{2015}).

\bibitem[{\citenamefont{Weisbuch and Stauffer}(2007)}]{weisbuch_pa07}
\bibinfo{author}{\bibfnamefont{G.}~\bibnamefont{Weisbuch}} \bibnamefont{and}
  \bibinfo{author}{\bibfnamefont{D.}~\bibnamefont{Stauffer}},
  \bibinfo{journal}{Physica A} \textbf{\bibinfo{volume}{384}},
  \bibinfo{pages}{542} (\bibinfo{year}{2007}).

\bibitem[{\citenamefont{Blume}(1993)}]{blume_l_geb93}
\bibinfo{author}{\bibfnamefont{L.~E.} \bibnamefont{Blume}},
  \bibinfo{journal}{Games Econ. Behav.} \textbf{\bibinfo{volume}{5}},
  \bibinfo{pages}{387} (\bibinfo{year}{1993}).

\bibitem[{\citenamefont{Nishimori}(2001)}]{nishimori_01}
\bibinfo{author}{\bibfnamefont{H.}~\bibnamefont{Nishimori}},
  \emph{\bibinfo{title}{{Statistical Physics of Spin Glasses and Information
  Processing: An Introduction}}} (\bibinfo{publisher}{Clarendon Press, Oxford,
  UK}, \bibinfo{year}{2001}).

\bibitem[{\citenamefont{Galam and Walliser}(2010)}]{galam_pa10}
\bibinfo{author}{\bibfnamefont{S.}~\bibnamefont{Galam}} \bibnamefont{and}
  \bibinfo{author}{\bibfnamefont{B.}~\bibnamefont{Walliser}},
  \bibinfo{journal}{Physica A} \textbf{\bibinfo{volume}{389}},
  \bibinfo{pages}{481} (\bibinfo{year}{2010}).

\bibitem[{\citenamefont{Matsuda et~al.}(1992)\citenamefont{Matsuda, Ogita,
  Sasaki, and Sato}}]{matsuda_h_ptp92}
\bibinfo{author}{\bibfnamefont{H.}~\bibnamefont{Matsuda}},
  \bibinfo{author}{\bibfnamefont{N.}~\bibnamefont{Ogita}},
  \bibinfo{author}{\bibfnamefont{A.}~\bibnamefont{Sasaki}}, \bibnamefont{and}
  \bibinfo{author}{\bibfnamefont{K.}~\bibnamefont{Sato}},
  \bibinfo{journal}{Progr. Theor. Phys.} \textbf{\bibinfo{volume}{88}},
  \bibinfo{pages}{1035} (\bibinfo{year}{1992}).

\bibitem[{\citenamefont{Schuster and Sigmund}(1983)}]{schuster_jtb83}
\bibinfo{author}{\bibfnamefont{P.}~\bibnamefont{Schuster}} \bibnamefont{and}
  \bibinfo{author}{\bibfnamefont{K.}~\bibnamefont{Sigmund}},
  \bibinfo{journal}{J. Theor. Biol.} \textbf{\bibinfo{volume}{100}},
  \bibinfo{pages}{533} (\bibinfo{year}{1983}).

\bibitem[{\citenamefont{Szolnoki and Perc}(2014)}]{szolnoki_pre14}
\bibinfo{author}{\bibfnamefont{A.}~\bibnamefont{Szolnoki}} \bibnamefont{and}
  \bibinfo{author}{\bibfnamefont{M.}~\bibnamefont{Perc}},
  \bibinfo{journal}{Phys. Rev. E} \textbf{\bibinfo{volume}{89}},
  \bibinfo{pages}{022804} (\bibinfo{year}{2014}).

\bibitem[{\citenamefont{Fort and Viola}(2005)}]{fort_jsm05}
\bibinfo{author}{\bibfnamefont{H.}~\bibnamefont{Fort}} \bibnamefont{and}
  \bibinfo{author}{\bibfnamefont{S.}~\bibnamefont{Viola}}, \bibinfo{journal}{J.
  Stat. Mech. Theor. Exp.} \textbf{\bibinfo{volume}{2}},
  \bibinfo{pages}{P01010} (\bibinfo{year}{2005}).

\bibitem[{\citenamefont{Vainstein and Arenzon}(2001)}]{vainstein_pre01}
\bibinfo{author}{\bibfnamefont{M.~H.} \bibnamefont{Vainstein}}
  \bibnamefont{and} \bibinfo{author}{\bibfnamefont{J.~J.}
  \bibnamefont{Arenzon}}, \bibinfo{journal}{Phys. Rev. E}
  \textbf{\bibinfo{volume}{64}}, \bibinfo{pages}{051905}
  (\bibinfo{year}{2001}).

\bibitem[{\citenamefont{Arapaki}(2009)}]{arapaki_pa09}
\bibinfo{author}{\bibfnamefont{E.}~\bibnamefont{Arapaki}},
  \bibinfo{journal}{Physica A} \textbf{\bibinfo{volume}{388}},
  \bibinfo{pages}{2757} (\bibinfo{year}{2009}).

\bibitem[{\citenamefont{Binder and Hermann}(1988)}]{binder_88}
\bibinfo{author}{\bibfnamefont{K.}~\bibnamefont{Binder}} \bibnamefont{and}
  \bibinfo{author}{\bibfnamefont{D.~K.} \bibnamefont{Hermann}},
  \emph{\bibinfo{title}{Monte Carlo Simulations in Statistical Physics}}
  (\bibinfo{publisher}{Springer}, \bibinfo{address}{Heidelberg},
  \bibinfo{year}{1988}).

\bibitem[{\citenamefont{Binder}(1997)}]{binder_rpp97}
\bibinfo{author}{\bibfnamefont{K.}~\bibnamefont{Binder}},
  \bibinfo{journal}{Rep. Prog. Phys.} \textbf{\bibinfo{volume}{60}},
  \bibinfo{pages}{487} (\bibinfo{year}{1997}).

\bibitem[{\citenamefont{Huberman and Glance}(1993)}]{huberman_pnas93}
\bibinfo{author}{\bibfnamefont{B.}~\bibnamefont{Huberman}} \bibnamefont{and}
  \bibinfo{author}{\bibfnamefont{N.}~\bibnamefont{Glance}},
  \bibinfo{journal}{Proc. Natl. Acad. Sci. USA} \textbf{\bibinfo{volume}{90}},
  \bibinfo{pages}{7716} (\bibinfo{year}{1993}).

\bibitem[{\citenamefont{Buonanno et~al.}(2009)\citenamefont{Buonanno, Montolio,
  and Vanin}}]{Buonanno2009}
\bibinfo{author}{\bibfnamefont{P.}~\bibnamefont{Buonanno}},
  \bibinfo{author}{\bibfnamefont{D.}~\bibnamefont{Montolio}}, \bibnamefont{and}
  \bibinfo{author}{\bibfnamefont{P.}~\bibnamefont{Vanin}}, \bibinfo{journal}{J.
  Law Econ.} \textbf{\bibinfo{volume}{52}}, \bibinfo{pages}{145}
  (\bibinfo{year}{2009}).

\bibitem[{\citenamefont{Botzen}(2016)}]{Botzen2016}
\bibinfo{author}{\bibfnamefont{K.}~\bibnamefont{Botzen}},
  \bibinfo{journal}{REGION} \textbf{\bibinfo{volume}{3}}, \bibinfo{pages}{1}
  (\bibinfo{year}{2016}).

\bibitem[{\citenamefont{Rand et~al.}(2014)\citenamefont{Rand, Peysakhovich,
  Kraft-Todd, Newman, Wurzbacher, Nowak, and Greene}}]{rand_nc14}
\bibinfo{author}{\bibfnamefont{D.~G.} \bibnamefont{Rand}},
  \bibinfo{author}{\bibfnamefont{A.}~\bibnamefont{Peysakhovich}},
  \bibinfo{author}{\bibfnamefont{G.~T.} \bibnamefont{Kraft-Todd}},
  \bibinfo{author}{\bibfnamefont{G.~E.} \bibnamefont{Newman}},
  \bibinfo{author}{\bibfnamefont{O.}~\bibnamefont{Wurzbacher}},
  \bibinfo{author}{\bibfnamefont{M.~A.} \bibnamefont{Nowak}}, \bibnamefont{and}
  \bibinfo{author}{\bibfnamefont{J.~D.} \bibnamefont{Greene}},
  \bibinfo{journal}{Nat. Commun.} \textbf{\bibinfo{volume}{5}},
  \bibinfo{pages}{3677} (\bibinfo{year}{2014}).

\bibitem[{\citenamefont{Capraro et~al.}(2014)\citenamefont{Capraro, Jordan, and
  Rand}}]{capraro2014heuristics}
\bibinfo{author}{\bibfnamefont{V.}~\bibnamefont{Capraro}},
  \bibinfo{author}{\bibfnamefont{J.~J.} \bibnamefont{Jordan}},
  \bibnamefont{and} \bibinfo{author}{\bibfnamefont{D.~G.} \bibnamefont{Rand}},
  \bibinfo{journal}{Sci. Rep.} \textbf{\bibinfo{volume}{4}},
  \bibinfo{pages}{6790} (\bibinfo{year}{2014}).

\bibitem[{\citenamefont{Capraro and Cococcioni}(2015)}]{capraro2015social}
\bibinfo{author}{\bibfnamefont{V.}~\bibnamefont{Capraro}} \bibnamefont{and}
  \bibinfo{author}{\bibfnamefont{G.}~\bibnamefont{Cococcioni}},
  \bibinfo{journal}{Proc. R. Soc. B} \textbf{\bibinfo{volume}{282}},
  \bibinfo{pages}{20150237} (\bibinfo{year}{2015}).

\bibitem[{\citenamefont{Bear and Rand}(2016)}]{bear_pnas16}
\bibinfo{author}{\bibfnamefont{A.}~\bibnamefont{Bear}} \bibnamefont{and}
  \bibinfo{author}{\bibfnamefont{D.~G.} \bibnamefont{Rand}},
  \bibinfo{journal}{Proc. Natl. Acad. Sci.} \textbf{\bibinfo{volume}{113}},
  \bibinfo{pages}{936} (\bibinfo{year}{2016}).

\end{thebibliography}
\end{document}